\documentclass{aa}  
\usepackage{graphicx}
\usepackage{txfonts}
\usepackage[figuresright]{rotating}
\usepackage{natbib}
\bibpunct{(}{)}{;}{a}{}{,}
\begin{document}

   \title{Polycyclic aromatic hydrocarbon processing in interstellar shocks}

   \author{E. R. Micelotta\inst{1,2}, A. P. Jones\inst{2}, A. G. G. M. Tielens\inst{1,3}}

   \offprints{E. R. Micelotta}

   \institute{Sterrewacht Leiden, Leiden University, P.O. Box 9513, 2300 RA 
              Leiden, The Netherlands\\
              \email{micelot@strw.leidenuniv.nl}
              \and
              Institut d'Astrophysique Spatiale, Universit\'{e} Paris Sud and CNRS (UMR 8617),
              91405 Orsay, France
              \and
              NASA Ames Research Center, MS 245-3, Moffett Field, CA 94035, USA\\ 
             }

   \date{Received 20 January 2009; accepted 16 October 2009}

  \abstract
    {PAHs appear to be an ubiquitous interstellar dust component but the effects of shocks
     waves upon them have never been fully investigated.}
   {We study the effects of energetic ($\approx 0.01-1$ keV) ion (H, He and C) and electron
    collisions on PAHs in interstellar shock waves.}
   {We calculate the ion-PAH and electron-PAH nuclear and electronic interactions, above the
    threshold for carbon atom loss from a PAH, in $50-200$ km s$^{-1}$ shock waves in the
    warm intercloud medium.}
   {Interstellar PAHs ($N_{\rm C} = 50$) do not survive in shocks with velocities greater
    than 100 km s$^{-1}$ and larger PAHs ($N_{\rm C} = 200$) are destroyed for shocks with
    velocities $\geq 125$ km s$^{-1}$. For shocks in the $\approx 75 - 100$ km s$^{-1}$ range, 
    where destruction is not complete, the PAH structure is likely to be severely denatured by 
    the loss of an important fraction ($20-40$\%) of the carbon atoms. We derive typical PAH 
    lifetimes of the order of a few $\times 10^8$ yr for the Galaxy. These results are robust 
    and independent of the uncertainties in some key parameters that have yet to be 
    well-determined experimentally.}
   {The observation of PAH emission in shock regions implies that that emission either arises
    outside the shocked region or that those regions entrain denser clumps that, unless they
    are completely ablated and eroded in the shocked gas, allow dust and PAHs to survive in
    extreme environments.}

   \authorrunning{E. R. Micelotta et al.}

   \titlerunning{PAH processing in interstellar shocks}
   \maketitle


\section{Introduction}

Interstellar Polycyclic Aromatic Hydrocarbon molecules (PAHs) are an ubiquitous component 
of the interstellar medium. The mid-infrared spectrum of the general diffuse interstellar 
medium as well as energetic environments near massive stars such as {\sc H\,ii} regions and 
reflection nebulae are dominated by broad emission features at 3.3, 6.2, 7.7, and 11.2 
$\mu$m. These emission features are now generally attributed to infrared fluorescence by 
large PAH molecules containing 50-100 C-atoms, pumped by single FUV photons \citep[see]
[ for a recent review]{tielens08}. The observed spectra also show evidence for PAH clusters 
containing a few hundred C-atoms \citep{breg89, rap05, berne07} as well as very small dust 
grains \citep[$\sim$30 \AA; ][]{desert90}. It seems that the interstellar grain size distribution 
extends all the way into the molecular domain \citep{allamandola89, desert90, draine01}. 
The origin and evolution of interstellar PAHs are somewhat controversial. On the one hand, 
based upon extensive laboratory studies of soot formation in terrestrial environments, 
detailed models have been made for the formation of PAHs in the ejecta of C-rich giants 
\citep{frenk89, cher92} -- as intermediaries or as side-products of the soot-formation 
process -- and studies have suggested that such objects might produce enough 
PAHs to seed the ISM \citep{latter91}. On the other hand, models have been developed where 
PAHs (as well as very small grains) are the byproduct of the grinding-down process of large 
carbonaceous grains in strong supernova shock waves which permeate the interstellar medium 
\citep{bork95, jones96}. Grain-grain collisions shatter fast moving dust grains into small 
fragments and, for graphitic progenitor grains, these fragments might be more properly 
considered PAH molecules. The destruction of interstellar PAHs is equally clouded. 
Laboratory studies have shown that small (less than 16 C-atoms), (catacondensed) PAHs are 
rapidly photodissociated by $\sim$ 10 eV photons \citep{joch94}. However, this process 
is strongly size-dependent as larger PAHs have many more modes over which the internal 
energy can be divided and PAHs as large as 50 C-atoms might actually be stable against 
photodissociation in the ISM \citep{lePage01, allamandola89}. While strong shock waves 
have been considered as formation sites for interstellar PAHs, the destruction 
of these PAHs in the hot postshock gas has not been evaluated. Yet, high energy ($\sim 1$ keV) 
collisions of PAHs with ions and electrons are highly destructive. 

The observational evidence for PAHs in shocked regions is quite ambiguous. The majority 
of supernova remnants does not show PAH features \citep[e.g. \object{Cas A},][]{smith08}, but 
observations of \object{N132D} \citep{tappe06} suggest the possibility of PAH survival 
in shocks. Recent work by \citet{and07} investigates the presence of PAHs in a 
subset of galactic supernova remnants detected in the GLIMPSE survey. Unfortunately 
the interpretation of such observations is not straightforward, because of the 
difficulty in disentangling the PAH features intrinsic to the shocked region with 
those arising from the surrounding material. Another interesting case is the starburst 
galaxy \object{M82}, which shows above and below the galactic plane a huge bipolar outflow of 
shock-heated gas interwoven with PAH emission\footnote{http://chandra.harvard.edu/photo/2006/m82/} 
\citep{armus07}. PAHs have also been observed at high galactic latitudes in the edge-on 
galaxies \object{NGC 5907} and \object{NGC 5529} \citep{irwin06, irwin07}. Shock driven winds and supernovae 
can create a so-called "galactic fountain" \citep{bregman80} transporting material into the 
halo and these detections of PAHs suggests the possibility of survival or formation of 
the molecules under those conditions. On the other hand \citet{ohalloran06, ohalloran08}
have found a strong anti-correlation between the ratio [FeII]/[NeII] and PAH strenght in
a sample of low-metallicity starburst galaxies. Since [FeII] has been linked primarly
to supernova shocks, the authors attributed the observed trend to an enhanced
supernova activity which led to PAH destruction.

In our previous study \citep{jones96}, we considered the dynamics and processing of
small carbon grains with $N_{\rm C} \geq$~100. The processing of these grains by
sputtering (inertial and thermal) in ion-grain collisions and by vaporisation and
shattering in grain-grain collisions was taken into account for all the considered grain
sizes. In that work, the smallest fragments ($a_{\rm f} < 5$\,\AA) were collected in the smallest
size bin and not processed. In this work we now consider what happens to these smallest
carbon grain fragments that we will here consider as PAHs. In this paper, we will consider 
relatively low velocity ($\leq 200$ km s$^{-1}$) shocks where the gas cools rapidly behind 
the shock front but, because of their inertia, PAHs (and grains) will have high velocity 
collisions even at large postshock column densities. Collisions between PAHs and the gas ions 
occur then at the PAH velocity which will slowly decrease behind the shock front due to the gas 
drag. This relative velocity is thus independent of the ion mass and, for dust grains, destruction 
is commonly called inertial sputtering. Destruction of PAHs in high velocity ($\geq 200$ km s$^{-1}$) 
shocks -- which cool slowly through adiabatic expansion -- is dominated by thermal sputtering and 
these shocks are considered in a subsequent paper \citep[][ hereafter MJT]{micelotta09}.

This paper is organized as follows:
Sect. 2 describes the theory of ion interaction with solids, Sect. 3 illustrates the
application of this theory to PAH processing by shocks and Sect. 4 presents our
results on PAH destruction. The PAH lifetime in shocks and the astrophysical 
implications are discussed in Sect. 5 and our conclusions summarized in Sect. 6.

\section{Ion interaction with solids}

\subsection{Nuclear interaction}

The approach used in our earlier work is not valid for planar PAH
molecules with of the order of tens of carbon atoms.  Here, we assume
that collisions are binary in nature, as is assumed in work on solids
\citep{lind63, lind68, sig81}.  If the energy transfer is above the
appropriate threshold value, we assume that the carbon target is
ejected from the molecule. For energy transfer below that threshold,
the energy will become thermal energy and be radiated away.

In this description the ``bulk'' nature of the target enters only
after the first interaction, when the projectile propagates into the
material. We therefore consider only the first interaction, which is
described in the binary collision approximation in a way that then
conveniently allows us to take into account the ``molecular'' nature
of the target.

In addition to the energy directly transferred to the target nucleus
through elastic scattering (\textit{nuclear stopping} or
\textit{elastic energy loss}), the energy loss to the atomic electrons
(\textit{electronic stopping} or \textit{inelastic energy loss})
should also be considered \citep{lind63, lind68}. In a solid the
energy transferred via electronic excitation is distributed around the
impact region. For a PAH, which has a finite size, the energy will be
spread out over the entire molecule. This energy will either be
radiated away or a fragment can be ejected.

Nuclear and electronic stopping are simultaneous processes which can be treated separately
\citep{lind63}. Fig. \ref{PAHevol_fig} illustrates these 
effects and shows the PAH evolution following the loss of carbon atoms, $N_{\rm C}$(lost), 
for the two limiting cases: 1) where there is an instantaneous and random removal of 
the lost carbon atoms and 2) where the carbon atoms are removed only from the periphery 
in order to preserve aromatic domain as much as possible. The reality of PAH erosion 
in shocks probably lies somewhere between these two extremes and will involve 
isomerisation and the formation of five-fold carbon rings that distort the structure 
from a perfectly two-dimensional form. This then begs the question as to the exact 
form and structure of small carbon species once growth resumes by atom insertion 
and addition. The full treatment of the nuclear stopping is given here, for the electronic 
stopping only the results of the calculations are shown, for the complete description 
of the phenomenon we refer the reader to paper MJT.

The treatment of PAH processing by shocks should also include the effects of 
fast electrons present in the gas. Because of their low mass, electrons can 
reach high velocities and hence high collision rates even at relatively low 
temperatures ($T \sim$10$^{5}$~K), leading to potentially destructive collisions. 
Again for a detailed description of the electron-PAH interaction see paper MJT. 

The theory of ion penetration into solids described here considers collisions
where the transferred energy $T$ goes from 0 to the maximum transferable energy. 
For this study, we are interested in only those collisions that are able to remove 
carbon atoms from the PAH, i.e. for which the energy transferred is greater than the 
minimum energy $T_{0}$ required for C ejection. In Sect. 2.2 we present the modifications
we introduce into the theory in order to treat the case of collisions above this threshold.  

To describe the binary collision between a moving atom (or ion) 
and a stationary target atom \citep[e.g.][]{sig81}, a pure classical two-particle
model using the Coulomb repulsion between the nuclei (Rutherford
scattering) is adequate only at high energies, i.e. when $\varepsilon\gg$~1, 
where $\varepsilon$ is the dimensionless \textit{Lindhard's reduced energy}

\begin{equation}\label{eps_eq}
 \varepsilon = \frac{M_{2}}{M_{1}+M_{2}}\;\frac{a}{Z_{1}\,Z_{2}\,e^{2}}\;E
\end{equation}
\smallskip
\noindent
where $M_{1}$ and  $Z_{1}$ are the mass and atomic number of incident particle 
respectively, $M_{2}$ and $Z_{2}$ the mass and atomic number of target particle,
$E$ is the kinetic energy of incident particle and $e$ is the electron charge, with
$e^{2}$=14.39 eV$\AA$. 
The quantity $a$ is the \textit{screening length}, a parameter that defines the
radial spread of the electronic charge about the nucleus.
For the screening length we adopt the Universal 
Ziegler - Biersack - Littmark (ZBL) screening length $a_{\rm U}$ 
\citep{zie85}
\begin{equation}\label{aU_eq}
 a_{\rm U} \cong 0.885\:a_{0}\:(Z^{0.23}_{1}+\,Z_{2}^{0.23})^{-1}
\end{equation}
\noindent
where $a_{0}$ = 0.529 $\AA$ is the Bohr radius.
The condition $\varepsilon$~$\gg$~1 implies that the energies are large enough that 
the nuclei approach closer to each other than the screening length $a$. 
At lower energies, $\varepsilon \lesssim$ 1, it is essential to consider the 
\textit{screening} of the Coulomb interaction.
In this case the Rutherford approximation is not adequate and 
the scattering problem must be treated using a different approach.

To choose the appropriate formalism to describe our interaction, we need 
to calculate the reduced energy for our projectiles. 
For our study of the behaviour of PAHs in shocks, we consider the binary
collision between H, He and C ions (projectiles) and a carbon atom (target) 
in the PAH molecule. The velocity $v_{\rm p}$ of the projectile is determined by 
the shock velocity $v_{\rm S}$ through the relation $v_{\rm p} = \frac{3}{4}\,$$v_{\rm S}$.
We consider here shock velocities between 50 and 200 km s$^{-1}$.
The corresponding projectile kinetic energies $E$ and reduced energies 
$\varepsilon$ are reported in Table \ref{energy_tab} for the two limiting cases 
$v_{\rm p1} = \frac{3}{4}$(50) = 37.5 km s$^{-1}$ and $v_{\rm p2} = \frac{3}{4}$(200) 
= 150 km s$^{-1}$.


\begin{figure}
  \centering
  \includegraphics[width=.9\hsize]{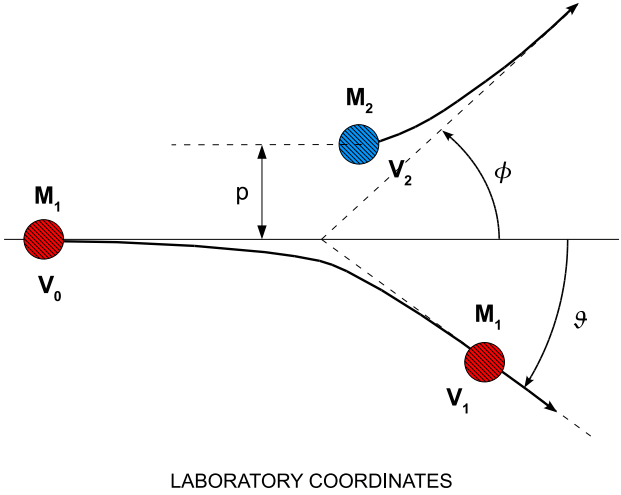}
  \caption{Scattering geometry for an elastic collision of particle 1
           (mass $M_{1}$, initial velocity $v_{0}$, impact parameter $p$), 
           on particle 2 (mass $M_{2}$, initial velocity zero). After
           the impact, the projectile particle 1 is deflected by the 
           angle $\vartheta$ and continues its trajectory with velocity 
           $v_{1}$. The target particle 2 recoils at an angle $\phi$
           with velocity $v_{2}$.
           }
  \label{Scattering_fig}
\end{figure}



\begin{table}
  \begin{minipage}[t]{\columnwidth}
    \caption
        {Kinetic energy $E$ and reduced energy $\varepsilon$ for H, He
          and C impacting on a carbon atom. 
        }
        \label{energy_tab}      
        \centering          
        \renewcommand{\footnoterule}{}      
        \begin{tabular}{c c c c c}     
          \noalign{\smallskip}
          \noalign{\smallskip}
          \hline\hline       
          \noalign{\smallskip}
          Projectile  &  $E$\footnote{Kinetic energy in eV. The projectile 
          velocities $v_{\rm p1}$ = 37.5 km s$^{-1}$ and
          $v_{\rm p2}$ = 150 km s$^{-1}$ are defined by the shock velocity $v_{\rm
            S}$ via the equation $v_{\rm p}$ = $\frac{3}{4}$ $v_{\rm S}$,
          with $v_{\rm S}$ = 50 and 200 km s$^{-1}$ respectively.}($v_{\rm p1}$)  &  
          $\varepsilon$\footnote{Dimensionless, calculated from Eq. \ref{eps_eq} and
          Eq. \ref{aU_eq}}($v_{\rm p1}$)  &  $E^{a}$($v_{\rm p2}$)  &  
          $\varepsilon^{b}$($v_{\rm p2}$) \\
          \noalign{\smallskip}
          \hline                    
          \noalign{\smallskip}
          H    &   7.30  &   0.031  &   117.4  &   0.50  \\
          He   &   29.4  &   0.048  &   469.7  &   0.76  \\
          C    &   88.1  &   0.028  &   1409.  &   0.45  \\
          \noalign{\smallskip}
          \hline 
        \end{tabular}
  \end{minipage}
\end{table}      


The calculation clearly shows that for the shocks we are considering
$\varepsilon \lesssim$ 1, implying that our problem cannot be treated
in terms of Rutherford scattering but requires a different formalism, 
described by \citet{sig81} and summarized below.
The scattering geometry for an elastic collision of
the projectile particle 1 on target particle 2 is illustrated in
Fig. \ref{Scattering_fig}.  Particle 1 has mass $M_{1}$, initial
velocity $v_{0}$ and impact parameter $p$, where the impact
parameter is the distance of closest approach of the centers of the
two atoms/ions that would result if the projectile trajectory was
undeflected. Particle 2 has mass $M_{2}$ and is initially at
rest. After the impact, the projectile is deflected by the angle 
$\vartheta$ and continues its trajectory with velocity $v_{1}$. A 
certain amount of energy $T$ is transferred to the target particle
which recoils at an angle $\phi$ with velocity $v_{2}$.
The maximum transferable energy corresponds to a head-on collision 
(impact parameter $p$ = 0) and is given by 
\begin{equation}\label{tm_eq}
 T_{\rm m} = \gamma\,E = \frac{4\,M_{1}\,M_{2}}{(M_{1}+M_{2})^{2}}\,E
\end{equation}

An important quantity to consider is the \textit{nuclear stopping cross section} 
$S_{\rm n}(E)$, which is related to the average energy loss per unit path length 
of a particle travelling through a material of atomic number density $N$ \citep{lind63}

\begin{equation}\label{dEdR_eq}
 \frac{{\rm d} E}{{\rm d} R} = N\,S_{\rm n}(E) = N\int {\rm d} \sigma(E,T) \cdot T
\end{equation}
\smallskip
\noindent
where $\sigma(E,T)$ is the energy transfer cross section (see Appendix A for details).
$S_{\rm n}(E)$ has the dimensions of (energy $\times$ area $\times$ atom$^{-1}$) and
in fact represents the average energy transferred per atom in elastic collisions
when summed over all impact parameters.

The nuclear stopping cross section can be expressed in terms of the 
Lindhard's reduced energy $\varepsilon$ and the dimensionless
\textit{reduced} nuclear stopping cross section $s_{\rm n}(\varepsilon)$ 
\citep[][ see Eq. \ref{Sn_gen_eq} and \ref{sn_red_eq}]{lind68}.
For this latter we adopt the Universal reduced Ziegler - Biersack - Littmark 
(ZBL) nuclear stopping cross section $s_{\rm n}^{\rm U}$ \citep{zie85}, which is an 
analytical approximation to a numerical solution that reproduces well the 
experimental data.
The ZBL reduced nuclear stopping cross section has the form

\begin{equation}\label{sn_ZBL_eq}
  s_{\rm n}^{\rm U}(\varepsilon) = \left\{ 
      \begin{array}{ll}
         \displaystyle \frac{0.5\,\ln\,(1+1.1383\,\varepsilon)}{\varepsilon+0.01321\,
           \varepsilon^{0.21226}+0.19593\,\varepsilon^{0.5}} & \;\;\; \varepsilon \le 30 \\[.5cm]
         \displaystyle \frac{\ln\,\varepsilon}{2\varepsilon} & \;\;\; \varepsilon > 30
      \end{array}
      \right.
\end{equation}
\smallskip
\noindent
and the nuclear stopping cross section $S_{\rm n}(E)$ can be written as

\begin{equation}\label{sn_totZBL_eq}
 S_{\rm n}(E) = 4\:\pi\:a_{\rm U}\:Z_{1}\:Z_{2}\:e^{2}\:\frac{M_{1}}{M_{1}+M_{2}}\;s_{\rm n}^{\rm U}(\varepsilon)
\end{equation}
\smallskip
\noindent
with the screening length $a_{\rm U}$ from Eq. \ref{aU_eq}.


\begin{table}
  \begin{minipage}[t]{\columnwidth}
    \caption
        {Threshold  energy $T_{0}$ and critical kinetic energy $E_{\rm 0n}$ for 
         H, He and C ions impacting on a carbon atom. 
        }
        \label{TH_tab}      
        \centering          
        \renewcommand{\footnoterule}{}      
        \begin{tabular}{c c c c}     
          \noalign{\smallskip}
          \noalign{\smallskip}
          \hline\hline       
          \noalign{\smallskip}
          $T_{0}$  &  $E_{\rm 0n}$(H)  &  $E_{\rm 0n}$(He)  &  $E_{\rm 0n}$(C)  \\
          \noalign{\smallskip}
          \hline                    
          \noalign{\smallskip}
             4.5    &   15.8  &  6.0  &   4.5  \\
             7.5    &   26.4  &  10.  &   7.5  \\
             10.    &   35.2  &  13.  &   10.  \\
             12.    &   42.3  &  16.  &   12.  \\
             15.    &   52.8  &  20.  &   15.  \\
          \noalign{\smallskip}
          \hline 
          \noalign{\smallskip}
          \noalign{\smallskip}
        \end{tabular}
  \end{minipage}
Note: $T_{0}$ and $E_{\rm 0n}$ in eV. \\
\end{table}      


\subsection{Nuclear interaction above threshold}

For this study we are interested in \textit{destructive} collisions, i.e., 
collisions for which the average transferred energy $T$ exceeds the minimum
energy $T_{0}$ required to remove a carbon atom from the PAH. The theory discussed
in Sect. 2.1 does not treat this situation and considers the specific case where
$T_{0}$~=~0 (no threshold). To include the treatment of collisions \textit{above} 
threshold ($T_{0}> $~0) we developped the appropriate expressions for the
relevant quantities described in the previous sections.

The definition of the nuclear stopping cross section $S_{\rm n}$(E) can be written
in a more general way as
\begin{equation}\label{SnTH_def_eq}
 S_{\rm n}(E) = \int_{T_{0}}^{T_{\rm m}}  {\rm d} \sigma(E,T)\cdot T
\end{equation}
where $T_{0} \geq$ 0. The \textit{total} energy transfer cross section per carbon atom 
$\sigma$(E) is defined by
\begin{equation}\label{sigmaTH_def_eq}
 \sigma(E) = \int_{T_{0}}^{T_{\rm m}}  {\rm d} \sigma(E,T)
\end{equation}
In this case the threshold $T_{0}$ must be strictly positive, otherwise $\sigma$ would
diverge (this can be verified by substituting the expression for d$\sigma$ from 
Eq. \ref{dsigma_eq} and evaluating the integral).
Finally, the \textit{average} energy transferred in  a binary collision 
is given by the ratio between $S_{\rm n}$ and $\sigma$
\begin{equation}\label{EtransfTH_def_eq}
 \langle T(E) \rangle = \frac{S_{\rm n}(E)}{\sigma(E)}
\end{equation}
\noindent

The condition $\gamma E = T_{\rm m} > T_{0} = \gamma\, E_{\rm 0n}$ then imposes, for the 
kinetic energy of the incoming ion, the condition that $E > E_{\rm 0n} = T_{0}/ \gamma$.
In Table \ref{TH_tab} we report the critical energies $E_{\rm 0n}$ for H, He and C
ions corresponding to different values of the threshold energy $T_{0}$.

Using d$\sigma$ from Eq. \ref{dsigma_eq}
and evaluating the above integrals we obtain
\begin{equation}\label{SnTH_eq}
 S_{\rm n}(E) = \frac{C_{m}\,E^{-m}}{1-m} \, [T_{\rm m}^{1-m} - T_{0}^{1-m}]
\end{equation}

\begin{equation}\label{sigmaTH_eq}
 \sigma(E) = \frac{C_{m}\,E^{-m}}{m} \, [T_{0}^{-m} - T_{\rm m}^{-m}]
\end{equation}

\begin{equation}\label{EtransfTH_eq}
 \langle T(E) \rangle = \frac{m}{1-m}\, \frac{T_{\rm m}^{1-m} - T_{0}^{1-m}}{T_{0}^{-m} - T_{\rm m}^{-m}}
\end{equation}

To calculate the quantity $m = m(E)$ we use the following expression 
from \citet{zie85}

\begin{equation}\label{m_eq}
  m(E) = 1 - \exp\,\left[\,-\exp \; \sum_{i=0}^{5}\,a_{i}\,\left(0.1\,\ln\,
    \left(\frac{\varepsilon(E)}{\varepsilon_{1}}\right)^{i}\,\right)\right]
\end{equation}
\noindent
with $\varepsilon_{1}$ = 10$^{-9}$ and $a_{i} =$ -2.432, -0.1509, 2.648, 
-2.742, 1.215, -0.1665.

Combining Eq. \ref{Sn_eq} and Eq. \ref{Sn_gen_eq}, after some
algebraic manipulation, we can rewrite the above expressions for
$S_{\rm n}$, $\sigma$ and $\langle T \rangle$ in the more convenient form shown
below. The full calculation is reported in Appendix A. As explained in
Sect. 2.1, we adopt for the reduced stopping cross section the ZBL
function $s_{\rm n}^{\rm U}(\varepsilon)$ (Eq. \ref{sn_ZBL_eq}) with the
appropriate screening length $a_{\rm U}$.

\begin{equation}\label{SnTH_use_eq}
  S_{\rm n}(E) = 4 \pi a Z_{1} Z_{2}\, e^{2} \frac{M_{1}}{M_{1}+M_{2}} s_{\rm n}^{\rm U}(\varepsilon)
  \left[1 - \left(\frac{E_{\rm 0n}}{E}\right)^{1-m}\right]
\end{equation}

\begin{equation}\label{sigmaTH_use_eq}
  \sigma(E) =  4 \pi a Z_{1} Z_{2} e^{2} \frac{M_{1}}{M_{1}+M_{2}} s_{\rm n}^{\rm U}(\varepsilon)
  \frac{1-m}{m}\frac{1}{\gamma\,E}
  \left[\left(\frac{E_{\rm 0n}}{E}\right)^{-m}-1\right]
\end{equation}

\begin{equation}\label{EtransfTH_use_eq}
  \langle T(E) \rangle = \frac{m}{1-m}\, \gamma \, \frac{E^{1-m} - E_{\rm 0n}^{1-m}}{E_{\rm 0n}^{-m} - E_{m}^{-m}}.
\end{equation}
\noindent
Note that the term outside of the square brackets in Eq. \ref{SnTH_use_eq} and 
\ref{sigmaTH_use_eq} is the stopping cross section $S_{\rm n}(E)$ when
$T_{0}$ = 0 (no threshold).

\begin{figure*}
  \centering
  \includegraphics[width=15.cm]{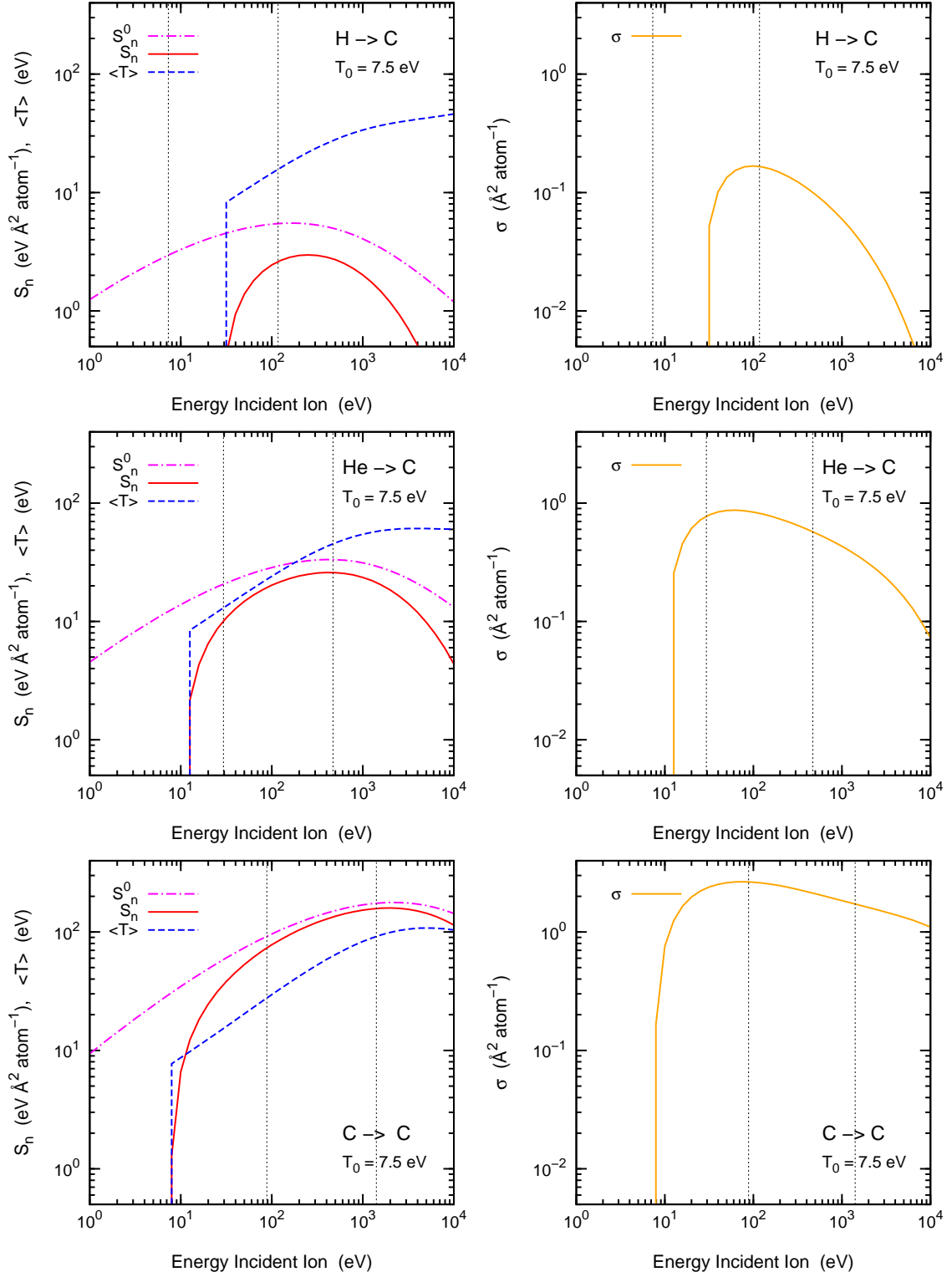}
  \caption{The nuclear stopping cross section $S_{\rm n}(E)$, the total cross
           section $\sigma(E)$ and the average energy transferred $\langle T(E) \rangle$
           calculated for H, He and C ions impacting on a carbon atom.
           The curves are calculated for the threshold energy $T_{0}$ = 7.5 eV.
           The nuclear stopping cross section $S^{0}_{\rm n}$ corresponding to $T_{0}$ = 0
           (no threshold) is shown for comparison. The two vertical lines 
           indicate the limiting energies for the incident ion. These are 
           defined as the kinetic energies of the projectile when its velocity 
           $v_{\rm p}$ equals $\frac{3}{4}$ ($v_{\rm S1,S2}$), where $v_{\rm S1}$ 
           = 50 km s$^{-1}$ and $v_{\rm S2}$ = 200 km s$^{-1}$ are the lowest and 
           highest shock velocities considered in this study.
           }
  \label{Stopping_ALL_fig}
\end{figure*}

The nuclear stopping cross section $S_{\rm n}(E)$ (eV $\AA^{2}$ atom$^{-1}$), 
the total energy transfer cross section $\sigma(E)$ ($\AA^{2}$ atom$^{-1}$)
and
the average energy transferred $\langle T(E) \rangle$ (eV), for H, He and C ions impacting
on a carbon atom, calculated from the above expressions assuming a threshold
$T_{0}$  = 7.5 eV, are shown in Fig.~\ref{Stopping_ALL_fig}.

The sharp cut on the left-hand side of the curves arises from the 
fact that we are treating collisions above threshold,
and these quantities 
are defined only for energies of the incident ion greater than the critical 
value $E_{\rm 0n}$. It can be seen that all
quantities increase in absolute value
with increasing atomic number and mass of the projectile ($Z_{1}$ and $M_{1}$).
The two vertical lines indicate the minimum and maximum kinetic energy of the 
projectile considered in our study, corresponding to the PAH velocity in 
the 50 and 200 km s$^{-1}$ shocks respectively. The values are those calculated 
in Sect. 2.1 and reported in Table \ref{energy_tab}. The figure cleary shows that
for hydrogen the critical value $E_{\rm 0n}$ is greater than the lower limit of energy 
range. This implies that in the lower velocity shocks hydrogen is not energetic 
enough to cause carbon ejection.
The curves for $S_{\rm n}$ presents a characteristic convex shape with a maximum, 
illustrating that nuclear energy transfer is important only for projectiles with 
energy falling in a specific range. In particular, the nuclear stopping becomes zero 
at high energies, with a limiting value depending on projectile and target: in our case, 
going from H to C impacting on carbon, the curves extend further to the right, in the 
direction of higher energies. In the high energy regime, the energy transfer is dominated by 
electronic stopping (MJT).

For a given incident ion energy, the difference between the values of $S_{\rm n}$
in the threshold and no-threshold cases results from the definition of the nuclear 
stopping and from the properties of d$\sigma$. The differential cross section
(cf. Eq. \ref{dsigma_eq}) strongly prefers collisions with low energy transfers
($T~\ll~T_{\rm m}$) and, moreover, decreases in absolute magnitute with increasing $E$.
For each $E$, $S_{\rm n}$ is defined as the integral over the transferred energy $T$, 
of the product between $T$ and the corresponding cross section d$\sigma$. 
Choosing $T_{0} > 0$ means excluding from the integral all energy transfers $T < T_{0} $,
for which the cross section has the highest values. The remaining terms have higher
values of $T$ but lower values of d$\sigma$, then the integral gives a result smaller 
than the no-threshold case, which includes all small energy transfers with their 
higher cross sections. 

The total cross section $\sigma$(E) clearly shows that the projectiles can efficiently 
transfer energy to the target atom only when their kinetic energy lies in the appropriate
window. In particular, it can be seen that the average energy transferred $\langle T \rangle$ increases 
with $E$, nevertheless at high energies $\sigma$ is close to zero (and the collision rate 
will be small). For a fixed target atom (in our case, carbon), the width of the $\sigma$ 
curve, and consequently the width of the energy window, increases with $Z$ and $M$ of the 
projectile. Heavier ions transfer more energy and in a more efficient way.\\
\clearpage


\clearpage
%
\noindent

\subsubsection{The threshold energy $T_{0}$}

The threshold energy, $T_{0}$, is the minimum energy that must be transferred via nuclear 
excitation to a carbon atom, in order to eject that same atom from the PAH molecule.
The choice for $T_{0}$ for a PAH is unfortunately not well-constrained. There are no 
experimental determinations, and the theoretical evaluation is uncertain.  
The analog of $T_{0}$ in a solid is the displacement energy $T_{\rm d}$, defined as the 
minimum energy that one atom in the lattice must receive in order to be moved more 
than one atomic spacing away from its initial position, to avoid the immediate hop 
back into the original site. For graphite, the data on the threshold energy for atomic 
displacement differ significantly, varying from $\sim$30~eV \citep{montet67, montet71} 
to 12 eV \citep{nakai91} largely depending on direction (eg., within or perpendicular 
to the basal plane). For a PAH, the lower value (corresponding to the perpendicular 
direction) seems then more appropriate. For amorphous carbon, \citet{coss78} has found 
a low value of 5 eV. Electron microscopy studies by \citet{ban97} on graphitic 
nanostructures irradiated with electrons of different energies, indicate that a value
of $T_{\rm d} \sim$~15-20 eV seems appropriate for the perpendicular direction. The in-plane 
value, however, could be much higher, presumably above 30 eV.

Instead of graphite, fullerenes and carbon nanotubes may be a better analog for PAH 
molecules. For fullerene, $T_{\rm d}$ has been found between 7.6 and 15.7 eV \citep{fuller96}.
Single walled nanotubes consist of a cylindrically curved graphene layer. Unfortunately, 
also in this case the threshold for atomic displacement is not precisely determined. However 
it is expected to be lower than in a multi-layered tube, for which a value of $T_{\rm d}~\sim$~15-20 eV 
has been found \citep{ban97} close to the value of graphite. We note that 4.5 and 7.5 eV 
are close to the energy of the single and double C-bond.

Because we cannot provide a well-defined $T_{0}$, we decided to explore a range 
of values, to study the impact of the threshold energy on the PAH processing.
For our standard case, we adopt 7.5 eV that we consider a reasonable value 
consistent with all the experimental data. However, we have varied $T_{0}$ 
from 4.5 to 15 eV.

\begin{figure}
  \begin{center}
   \includegraphics[width=.8\hsize]{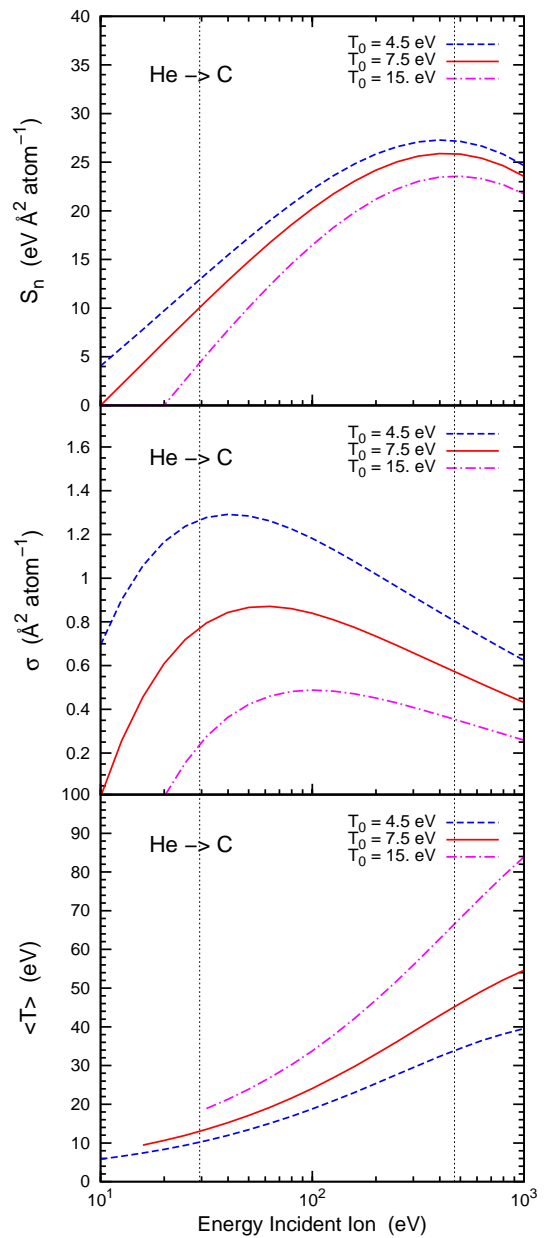}
  \end{center}
  \caption{The nuclear stopping cross section $S_{\rm n}(E)$, the cross
           section $\sigma(E)$ and the average energy transferred $\langle T(E) \rangle$
           calculated for He ions impacting on a carbon atom, calculated 
           for three values of the threshold energy $T_{0}$: 4.5, 7.5 and 15 eV.
           The two vertical lines delimit the energy range of interest
           (cf. Fig. \ref{Stopping_ALL_fig}).
           }
  \label{Stopping_He_fig}
\end{figure}

Fig. \ref{Stopping_He_fig} shows the comparison between $S_{\rm n}$, $\sigma$ and $\langle T \rangle$
calculated for He on C assuming $T_{0}$ = 4.5, 7.5 and 15 eV. Coherently with their 
definition, $S_{\rm n}$ and $\sigma$ increase with decreasing threshold, because more collisions 
are effective and the cross section increases with decreasing energy. Of course, the average 
energy transferred will decrease when the threshold energy is decreased. 

In Sect. 4.1 we discuss the effect of the choice of different values for $T_{0}$ on 
the PAH survival in shocks. 


\begin{figure}
  \centering
  \includegraphics[width=\hsize]{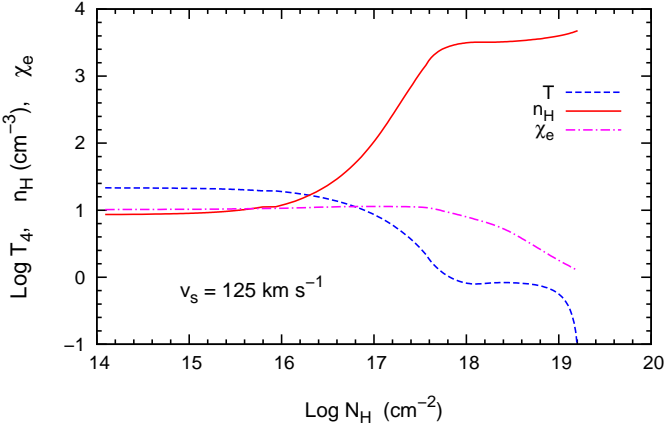}
  \caption{The structure of the 125 km s$^{-1}$ shock: temperature $T_{4} = T/$10$^{4}$~K,
           hydrogen density $n_{\rm H}$ and electron relative abundance $\chi_{\rm e}$.
           All quantities are plotted as a function of the shocked column
           density $N_{\rm H} = n_{0}\,$v$_{\rm S}\,t$, where the preshock density
           $n_{0}$ = 0.25 cm$^{-3}$. To convert column density to time, use the 
           following relation: log $t$(yr) = log $N_{\rm H}$(cm$^{-2}$) - 13.9
           \citep{jones96}.
           }
  \label{structureS_fig}
\end{figure}



\begin{figure}
  \centering
  \includegraphics[width=\hsize]{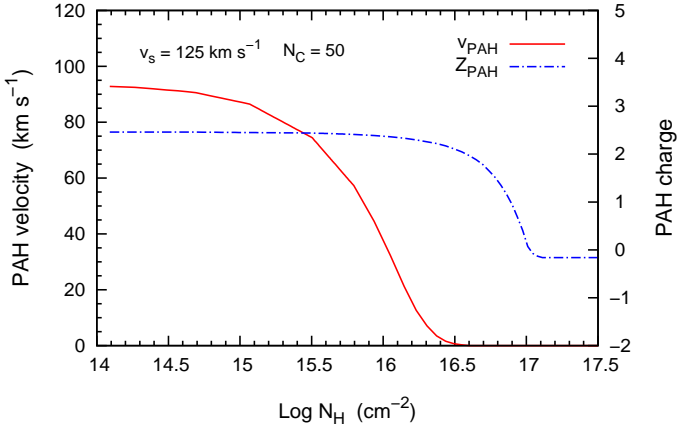}
  \caption{The velocity profile of a 50 C-atom PAH in a shock with
           velocity $v_{\rm S}$ = 125 km s$^{-1}$. The PAH velocity $v_{\rm PAH}$, 
           plotted as a function of the shocked column density $N_{\rm H}$, 
           represents the relative velocity between the molecule and the 
           ions present in the shock. Overlaid is the 
           PAH charge $Z_{\rm PAH}$ along the shock, calculated
           using the theory described in \citet{mckee87}.
           }
  \label{vProfile_fig}
\end{figure}



\begin{figure}
  \centering
  \includegraphics[width=\hsize]{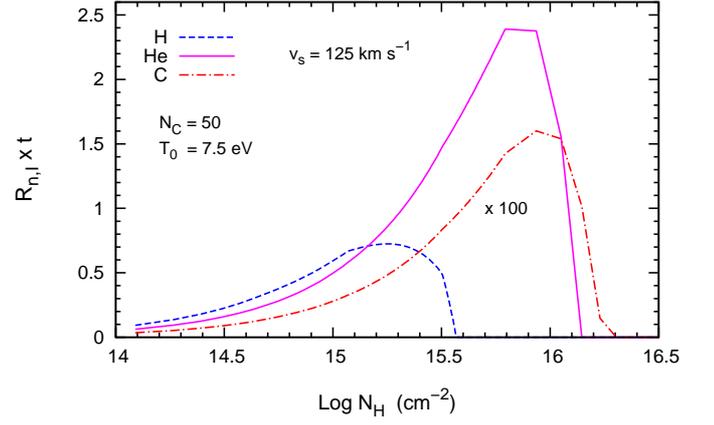}
  \caption{The number of collisions $N_{\rm t} = R_{\rm n,I} \times t$ of a 
           50 C-atom PAH with H, He and C ions in a 125 km s$^{-1}$ shock, 
           as a function of the shocked column density, $N_{\rm H}$. 
           The cross section has been evaluated for a threshold energy, 
           $T_{0}$, of 7.5 eV. The carbon curve has been multiplyied
           by a factor 100 for comparison.
           }
  \label{collTime_fig}
\end{figure}


\section{PAHs in shocks}

When grains and PAHs enter a shock they become charged and then gyrate around the
compressed magnetic field lines. This leads to relative gas-particle velocities and hence
to collisions with the gas (and other grains/PAHs). Collisions with the gas result in drag
forces and therefore a decrease in the relative gas-particle velocity. However, these same 
collisions with the gas can also lead to the removal of atoms from the particle if the 
relative velocites are larger than the given threshold for an erosional process.
The removal of carbon atoms from the PAH due to ion collisions, where the impact 
velocity is determined by the relative motion between the two partners, is the 
analog of the inertial sputtering of dust particles due to ion-grain collisions.
In the following we will then refer to it using the term \textit{inertial}, and 
the same will apply for all the related quantities.

In determining the processing of PAHs in shock waves, as with all grain processing, it is
the relative gas-grain velocity profile through the shock that determines the level of
processing. In calculating the relative ion-PAH velocity through the shock we use the 
same approach as in our previous work \citep{jones94, jones96}, which is based
on the methods described in \citet{mckee87}. The PAH velocity is calculated using a
3D particle of the same mass as the 50 carbon atom PAH under consideration. The PAH velocity 
depends then on the PAH mass and average geometric cross section. For a PAH with $N_{\rm C}$ carbon 
atoms, these are given by $N_{\rm C} m_{\rm C}$ and $0.5 \pi a_{\rm PAH}^2$ with $a_{\rm PAH}$ given by 
$0.9 \sqrt{N_{\rm C}}$ \AA , appropriate for a compact PAH \citep{omont86} and the factor $1/2$ in the cross 
section takes the averaging over impact angle into account.
The PAH and grain cross sections are very close (to within 11\% for $N_{\rm C}$ = 50), thus 
we are justified in using the same numerical approach even though we are using a 3D grain 
to calculate the velocity profile of a 2D PAH through the shock.

The PAHs are injected into the shock with 3/4 of the shock speed, as are all grains, and
their trajectories are then calculated self-consistently with their coupling to the
gas, until the relative gas-PAH veocity becomes zero. The velocity  calculation
includes the effects of the direct drag with the gas due to atom and ion collisions and
the drag due to the ion-charged PAH interaction in the post-shock plasma. We find that
for some shock velocities, in our case for $v_{\rm S} = 75$ and 100\,km\,s$^{-1}$,
the PAHs (and grains) experience betatron acceleration in the post-shock gas. All the
relevant expressions and assumptions for the calculation of the grain velocity, betatron
acceleration and grain charge are fully described in \citet{mckee87}. Thus, in
calculating the post-shock PAH velocity profiles, we follow exactly the same methods as
used in our previous work. The structure of the 125\,km\,s$^{-1}$ shock is shown in
Fig. \ref{structureS_fig} as a function of the column density $N_{\rm H}$. Fig.\ref{vProfile_fig} 
shows the velocity profile for a 50 carbon atom PAH in the same shock, together with the
effective charge of the molecule, used to calculate the velocity profile itself. The 50 
C-atoms PAH is positively charged (charge between $+$2 and $+$3) during the whole slowing 
process, and approaches neutrality at the end of the shock.

\subsection{Ion collisions: nuclear interaction}

Knowing the velocity profile of the PAH, we can then calculate the inertial 
collision rate PAH-ions $R_{\rm n,I}$ (s$^{-1}$) through the shock. This is 
given by the following equation
\begin{equation}\label{collRate_eq}
  R_{\rm n,I}(N_{\rm H}) = 0.5\;\chi_{i}\;n_{\rm H}\;v_{\rm PAH}\;\sigma\;N_{\rm C}\;F_{\rm C}
\end{equation}
where $n_{\rm H}(N_{\rm H})$ is the hydrogen particle density along the shock, 
v$_{\rm PAH}(N_{\rm H})$ the PAH-ion relative velocity along the shock and 
$\chi_{i}$ is the relative abundance of the projectile ion with respect to
hydrogen. We adopt the gas phase abundances $\chi_{\rm H}$~:~$\chi_{\rm He}$~:
~$\chi_{\rm C}$ = 1~:~10$^{-1}$~:~10$^{-4}$, where the carbon abundance is
between the values (0.5 - 1)$\times$10$^{-4}$ and 1.4$\times$10$^{-4}$
from \citet{sofia09} and \citet{card96} respectively.

The term $\sigma(N_{\rm H})$ is the cross section averaged over those collisions 
that transfer an energy larger than the threshold energy $T_{0}$ per C-atom and 
this cross section should therefore be multiplied by the number of carbon atoms 
in the PAH, $N_{\rm C}$. The factor 0.5 takes the angle averaged orientation into 
account (see appendix C).

Because both collision partners are charged, the effect of the Coulombian potential 
must be included as well. Depending on whether the interaction is attractive or
repulsive, the energy transfer cross section will be increased or reduced by the 
coulombian factor $F_{\rm C}$ given by
\begin{equation}
  F_{\rm C} = 1 \:-\: \frac{2\,Z_{\rm ion}\,Z_{\rm PAH}\,e^{2}}{4\,\pi\,
                  \varepsilon_{0}\,a_{\rm PAH}\,M_{1}\,m_{\rm H}\,v_{\rm PAH}^{2}}.
\end{equation}
where $Z_{\rm ion}$ = $+$1 and $M_{1}$ are the charge and the atomic mass (in amu) of the 
incident ion, $Z_{\rm PAH}$ is the charge of the PAHs along the shock, 
$a_{\rm PAH}$ is the PAH radius and v$_{\rm PAH}$ the PAH-ion relative velocity. 
The constant $e$ is the electron charge, $\varepsilon_{0}$ is the permittivity
of the free space and $m_{\rm H}$ the mass of the proton.
The total number of destructive collisions is then given by the integral of the collision 
rate (Eq. \ref{collRate_eq}) behind the shock,
\begin{equation}\label{Nt_eq}
N_{\rm t}\, =\, \int R_{\rm n,I}\left(N_{\rm H}\right)\, {\rm d}t
\end{equation}
where it should be understood that the postshock column density $N_{\rm H}$ and the time, $t$, are 
related through $N_{\rm H} = n_{0}v_{\rm S} t$. With the proper cross section, this number $N_{\rm t}$, 
is then equal to the number of carbon atoms lost by a PAH in collisions with H, He, or C.

Figure \ref{collTime_fig} illustrates the destructive collisions for a 50 C-atom PAH behind a 
125 km s$^{-1}$ shock assuming $T_0=7.5$ eV. These results are plotted in such a way that equal 
areas under the curve indicate equal contributions to the total number of destructive collisions.  
$R_{\rm n,I}$ drops precipitously because of the drop in relative PAH-gas 
velocity. Because heavier projectiles are more energetic in inertial collisions, this drop off 
shifts to higher column densities for heavier species. The results show that He is much more 
effective in destroying PAHs than H because of the increased energy transferred for heavier 
collision partners (cf. Figure \ref{Stopping_ALL_fig}). The low abundance of C depresses its 
importance in inertial sputtering. 

The number of carbon atoms in a PAH is now given by
\begin{equation}
  N_{\rm C}\left(t\right)\, =\,N_{\rm C}\left(0\right)\, \exp\left[ -N_{\rm t}/N_{\rm C}\right]
\end{equation}
and the fraction of carbon atoms ejected from this PAH is
\begin{equation}\label{frac_eq}
  F_{\rm L}\, =\, \left(1\, -\,  \exp\left[ -N_{\rm t}/N_{\rm C}\right]\, \right)
\end{equation}
where $N_{\rm t}$ is now evaluated throughout the shock.

In the shocked gas, the velocity of the ions is not only determined by the relative
motion with respect to the PAH (inertial case), but also by the temperature of
the shocked gas. In principle, the inertial and thermal velocity should be added vectorally 
and averaged over the angle between the inertial motion and the (random) thermal motion as 
well as over the thermal velocity distribution. However, that becomes a quite cumbersome 
calculation and, hence, we will follow calculations for sputtering of dust grains in 
interstellar shocks \citep[cf.][]{jones94} and evaluate these two processes (inertial and 
thermal sputtering) independently. Studies have shown that this reproduces more extensive 
calculations satisfactorily \citep{guillet07}. The thermal destruction rate is given by 
\begin{equation}\label{collRateThermal_eq}
  R_{\rm n,T}(N_{\rm H}) = N_{\rm C}\,0.5\,\chi_{i}\,n_{\rm H}\, \int_{v_{0}}^{\infty}
                         F_{\rm C}(v)\,v\,\sigma(v)\,f(v,T)\,{\rm d}v
\end{equation}
with $f\left(v, T\right)$ the Maxwellian velocity distribution. The temperature has to be 
evaluated along the shock profile (cf. Figure \ref{structureS_fig}) and care should be taken 
to only include velocities corresponding to energies larger than the threshold energy, $T_{0}$ 
(e.g., with $E>E_{\rm 0n}$; cf., Table \ref{TH_tab}). The fraction of C-atoms ejected by this 
process can be evaluated analogously to Eq. \ref{frac_eq}.

\subsection{Ion collisions: electronic interaction}

As reported in the introduction of the paper, the collision between PAH and ions
triggers two simultaneous process, which can be treated separately: the nuclear
stopping (\textit{elastic energy loss}) and the electronic stopping (\textit{inelastic 
energy loss}). The first has been extensively discussed in the previous sections, while
for the full treatment of the electronic interaction we refer the reader to MJT. For the 
sake of clarity, we report here the essential concepts and the principal equations which
will be used in the following.

The energy transferred to the electrons is spread out over the entire molecule,
leaving the PAH in an excited state. De-excitation occurs through two pricipal decay
channels: emission of infrared photons and dissociation and loss of a C$_{2}$ fragment. 
This latter is the process we are interested in, because it leads to the PAH fragmentation. 
The dissociation probability $p$ (see Sect. 4.1 in MJT) depends on the binding energy of the fragment 
$E_{0}$, on the PAH size, $N_{\rm C}$, and on the energy transferred, which in turns 
depends on the initial energy (velocity) of the projectile. 

For a fixed value of the transferred energy, the dissociation probability decreases 
for increasing $E_{0}$
and $N_{\rm C}$ because either more energy is required in the bond that has to be broken 
or because the energy is spread over more vibrational modes and hence the internal excitation 
temperature is lower. On the other hand, the more energy that is deposited in the PAH, the higher 
is the dissociation probability. The energy transferred via electronic excitation increases 
with the energy of the projectile up to a maximum value, corresponding to an incident energy 
of $~$100 keV for H (and higher for more massive particles), and decreases for 
higher energies. The deposited energy also increases with the path-length through the molecule
and will be higher for larger PAHs impacted at grazing collision angles. For the shocks 
considered in this study, the energy transferred increases with incident energy (velocity) 
and hence the dissociation probability increases as well.

As for the nuclear stopping, also for the electronic interaction we have to consider
the effect of both inertial and thermal velocities. The \textit{inertial} collision
rate is given by
\begin{equation}\label{collRateElec_eq}
  R_{\rm e,I}(N_{\rm H}) = v_{\rm PAH}\,\chi_{i}\,n_{\rm H}\,F_{\rm C}
                         \int_{\vartheta=0}^{\pi/2}\sigma_{\rm g}(\vartheta)\,
                         p(v_{\rm PAH},\vartheta)\,\sin\vartheta\,{\rm d}\vartheta
\end{equation}
where $\vartheta$ is the angle between the axis normal to the PAH plane and the direction
of the incoming ion. The term $\sigma_{\rm g}$ is the geometrical cross section seen by 
an incident particle with direction defined by $\vartheta$. The PAH is modelled as a thick 
disk with radius $a_{\rm PAH}$ and thickness $d$, then the cross section is given by
\begin{equation}\label{disk_cs_eq}
  \sigma_{\rm g} = \pi\,a_{\rm PAH}^{2}\,\cos\vartheta\,+\,2\,a_{\rm PAH}\,d\,\sin\vartheta
\end{equation}
which reduces to $\sigma_{\rm g} = \pi a_{\rm PAH}^{2}$ for $\vartheta$ = 0 (face-on impact) and to
$\sigma_{\rm g} = 2a_{\rm PAH}d$ for $\vartheta = \pi/2$ (edge-on impact).
The term $p(v_{\rm PAH},\vartheta)$ represents the total probability for dissociation
upon collision via electronic excitation, for a particle with relative velocity
$v_{\rm PAH}$ and incoming direction $\vartheta$ (see Sect. 4.1 in MJT).

For the \textit{thermal} collision rate we have
\begin{equation}\label{collRateElecThermal_eq}
 R_{\rm e,T}(N_{\rm H}) = \int_{v_{0}}^{\infty}R_{\rm e,I}(v)\,f(v,T)\,{\rm d}v
\end{equation}
where the temperature $T$ = $T(N_{\rm H})$ is evaluated along the shock. The lower
integration limit $v_{0}$ is the ion velocity corresponding to $E_{0}$. The number 
of carbon atoms lost can be evaluated analogously to the nuclear interaction but care 
should be taken to include the loss of 2 C-atoms per collision.

There is a clear distinction between the nuclear and electronic interactions. In nuclear 
interactions, a C-atom is ejected because a direct collision with the impacting ion 
transfers enough energy and momentum to kick out the impactee instantaneously. In 
electronic interaction, the impacting ion excites the electrons of the PAH. Internal 
conversion transfers this energy to the vibrational motions of the atoms of the PAH. 
Rapid intramolecular vibrational relaxation leads then to a thermalization of this excess 
energy among all the vibrational modes and this can ultimately lead to dissociation (or 
relaxation through IR emission). The threshold energy in the nuclear process, $T_{0}$, 
differs therefore from the electronic dissociation energy, $E_{0}$. The latter really 
is a parameter describing the dissociation rate of a highly excited PAH molecules using an 
Arrhenius law and this does not necessarily reflect the actual binding energy of the 
fragment to the PAH species \citep[cf.][]{tielens05}. Following MJT, we will adopt the 
canonical value of 4.6 eV for $E_{0}$. However, this energy is very uncertain and we will 
evaluate the effects of reducing and increasing this parameter to a value of 3.65 and 
5.6 eV respectively (MJT).

\subsection{Electron collisions}

For the full treatment of the PAH collisions with \textit{electrons}, we refer
again to the paper MJT, providing here a short summary of the basic concepts
and equations.

Because of their small mass, the thermal velocity of the electrons always exceeds the 
inertial velocity of the PAH. Hence, only the thermal destruction needs to be evaluated.
We follow the same formalism used for the electronic interaction in ion-PAH collisions.
The energy dumped into the molecule during collisions with electrons is spread over
and determines (with $E_{0}$ and $N_{\rm C}$) the value of the dissociation probability. 
The electron energy loss rises sharply with the electron energy, reaching its maximum
for incident energy around 100 eV. This energy range falls exactly in the interval 
relevant for our shocks, implying that the electrons optimally transfer their energy.

The thermal \textit{electron} collision rate can be written as
\begin{equation}\label{collRateELECTRON_eq}
 R_{\rm elec, T}(N_{\rm H}) = \int_{v_{\rm 0,elec}}^{\infty}\Sigma(v)\,f(v,T)\,{\rm d}v
\end{equation}
\noindent
\begin{equation}\label{SigmaELECTRON_eq}
  \Sigma(v) = v\,\chi_{\rm e}\,n_{\rm H}\,F_{\rm C e}
                         \int_{\vartheta=0}^{\pi/2}\sigma_{\rm g}(\vartheta)\,
                         p(v,\vartheta)\,sin\vartheta\,{\rm d}\vartheta
\end{equation}
where $v_{\rm 0, elec}$ is the electron velocity corresponding to $E_{0}$ and $\chi_{\rm e}$
is the electron relative abundance along the shock. The electron coulombian factor 
$F_{\rm C e}$ is always equal to 1 (within less than 1\%) because electrons have 
low mass and high velocities with respect to ions. The temperature $T$ evaluated
along the shock is the same as for the ions, but electrons will reach much larger
velocities. From Eq. \ref{collRateELECTRON_eq} and \ref{SigmaELECTRON_eq} we expect
then to find a significantly higher collision rate with respect to the ion case.  
The fraction of C-atoms lost by electron collisions can be evaluated analogously to 
that for ions (cf. Eq. \ref{frac_eq}).

\section{Results}

Fig. \ref{global50_fig} and \ref{global200_fig} show the fraction of carbon 
atoms ejected from a 50 and 200 C-atoms PAH due to collisions with electrons 
and H, He and C, assuming the nuclear threshold energy $T_{0}$~=~7.5~eV
and the fragment binding energy $E_{0}$~=~4.58~eV. The results concerning
nuclear, electronic and electron interaction are discussed in the following 
sections.

\subsection{PAH destruction via nuclear interactions}

For the \textit{inertial} nuclear interactions, the fraction of ejected carbon
atoms $F_{\rm L}$ depends on both $\sigma$ and $\chi$.
Hydrogen has the highest abundance ($\chi_{\rm H}$ = 1) 
but the lowest absolute value for the cross section (see Fig. \ref{Stopping_ALL_fig}). 
In addition, $\sigma$ is significantly different from zero only for the highest 
shock velocities. This results in contribution to atom ejection which is only 
relevant for $v_{\rm S}$ above 150 km s$^{-1}$. Helium is ten times less abundant 
than hydrogen ($\chi_{\rm He}$ = 0.1), but this is compensated for by a higher cross 
section for all shock velocities. In particular, the 
C-atom ejection curve shows a peak between 50 and 125 km/s due to betatron 
acceleration: because of the higher velocity, the collision rate increases 
(cf. Eq. \ref{collRate_eq}) and then the PAHs experience more destructive 
collisions. After the peak, as expected the curve increases with the shock 
velocity. In the case of carbon, the increased cross section is not sufficient 
to compensate for the low abundance ($\chi_{\rm C}$ = 10$^{-4}$), resulting in 
a totally negligible contribution to PAH destruction. For all shock velocities,
the fraction of C-atoms removed because of inertial nuclear interaction 
does not exceed the value of 20\%.

Concerning the \textit{thermal} nuclear interaction, carbon does not contribute 
to PAH destruction because of its very low abundance compared to H and He, 
as for the inertial case. For hydrogen and helium, as expected for low
velocity shocks, the temperature is generally not sufficiently high to provide 
the ions with the energy required to remove C-atoms. Nevertheless, the ions 
in the high velocity tail of the Maxwellian distribution can be energetic 
enough to cause C-atom ejection, as can be seen for He at 100 km s$^{-1}$.
This is less evident for hydrogen. In this case the critical energy $E_{\rm 0n}$
is higher than for helium and carbon. The corresponding critical velocity $v_{0}$ 
will be higher as well. For the lower velocity shocks, the peak of the hydrogen 
maxwellian function $f(v, T)$ is well below $v_{0}$, as a consequence the integrand 
of Eq. \ref{collRateThermal_eq} is close to zero over the integration range, and 
the same will be true for the collision rate. At the highest shock velocities the 
curves show a similar trend, with a steep rise beyond 125 km s$^{-1}$ leading 
to complete PAH destruction, i.e. removal of ALL carbon atoms, for shock 
velocities above 150 km s$^{-1}$. At around 135 km s$^{-1}$ the hydrogen contribution
becomes larger than that for helium. At these high velocities the He and H cross sections  
reach approximately their maximum values (cf. Fig. \ref{Stopping_ALL_fig}) and the 
abundance of H is a factor of 10 higher than for He.

As discussed in Sect. 2.2.1, the threshold energy for carbon ejection via nuclear 
excitation is not well-constrained. We consider $T_{0}$ = 7.5 eV to be a 
reasonable value, but experimental determinations are necessary. Fig. \ref{Ncoll3_fig} 
illustrates how the fraction of ejected C-atoms changes as a function of the 
adopted value
for the threshold energy. The curves show the cumulative effect 
of H, He and C, calculated for $T_{0}$ = 4.5, 7.5 and 15 eV in the inertial
and thermal case. Both in the inertial and thermal case, the curves corresponding 
to the various thresholds follow the same trend, and for each value considered of 
$T_{0}$ the inertial destruction dominates at low velocity and the thermal 
destruction at high velocities. As expected the fraction of ejected C-atoms increases 
for decreasing $T_{0}$ in the inertial case, while the curves shift to the left 
in the thermal case, implying that the PAHs will start to experience significant
damage at lower shock velocities. Our results also show that, even assuming a 
high threshold energy, PAHs experience a substantial loss of carbon atoms, which
is complete for velocities above 175 km s$^{-1}$ is all cases.

Finally, we investigated how the nuclear destruction process depends on the size 
of the PAH. Fig. \ref{global200_fig} shows the fraction of ejected carbon atoms 
from a big PAH with $N_{\rm C}$ = 200. The destruction of a 200 C-atom PAH follows 
the same trends with shock velocity as for the 50 C-atom case and the curves are 
almost identical. This is due to the fact that the velocity and temperature profiles 
for the 50 and 200 C-atoms molecules are quite similar, and the collision rate and 
$F_{\rm L}$ scale linearly with $N_{\rm C}$ in both the inertial and thermal case 
(see Eq. \ref{collRate_eq}, \ref{frac_eq} and \ref{collRateThermal_eq}).

\subsection{PAH destruction via electronic interaction by ion collisions}

Inspection of Fig. \ref{global50_fig} and \ref{global200_fig} reveals
that electronic excitation by impacting ions plays only a marginal role in 
the destruction process. For both PAH sizes, carbon is unimportant because 
of its very low abundance. In the inertial case H does not contribute and He 
contributes marginally at the highest shock velocities, while in the thermal 
case they lead to a substantial atomic loss only for $N_{\rm C}$~=~50 in the 
highest velocity shock (200 km s$^{-1}$). For a 50 C-atom PAH, the low destruction 
rate due to electronic excitation reflects the small cross section for this process 
for these low velocity shocks. The inertial velocities of the PAH lead to electronic 
excitation only being important for the highest shock velocities where the 
impacting ions have a high enough temperature to excite the PAHs sufficiently 
(cf. MJT). The larger number of modes available in 200 C-atom PAHs, makes the 
electronic excitation of such PAHs completely negligible over the full velocity 
range of the shocks considered here.

\subsection{PAH destruction due to electron collisions}

The fractional carbon atom loss $F_{\rm L}$ due to collisions with thermal electrons is also
shown in Fig. \ref{global50_fig} and \ref{global200_fig}. For $N_{\rm C}$ = 50, 
the number of ejected carbon atoms rises sharply above 75 km s$^{-1}$, leading 
to total destruction above 100 km s$^{-1}$. For $N_{\rm C}$ = 200, the damage is
negligible up to 100 km s$^{-1}$, increases significantly beyond that and leads to
complete destruction above 150~km~s$^{-1}$.

The energy transferred by impacting electrons rises sharply for velocities in excess 
of 2$\times$10$^{3}$ km s$^{-1}$. This results in a dissociation probability $p$ shaped 
as a step function: for $v \gtrsim$ 2$\times$10$^{3}$ km s$^{-1}$ $p$ jumps from values 
close to zero up to 1. This limiting velocity applies to a 50 C-atoms PAH; for 
$N_{\rm C}$ = 200 the value is higher (4$\times$10$^{3}$ km s$^{-1}$), due to the fact that 
for a bigger PAH  more energy has to be transferred for dissociation. These velocities 
correspond to electron temperatures of 10$^{5}$ K and 3$\times$10$^{5}$ K, which are 
reached for shock velocities of approximately 100 and 150 km s$^{-1}$, respectively.

\subsection{Summary}

\begin{figure*}[htp]
  \centering
  \includegraphics[width=.8\hsize]{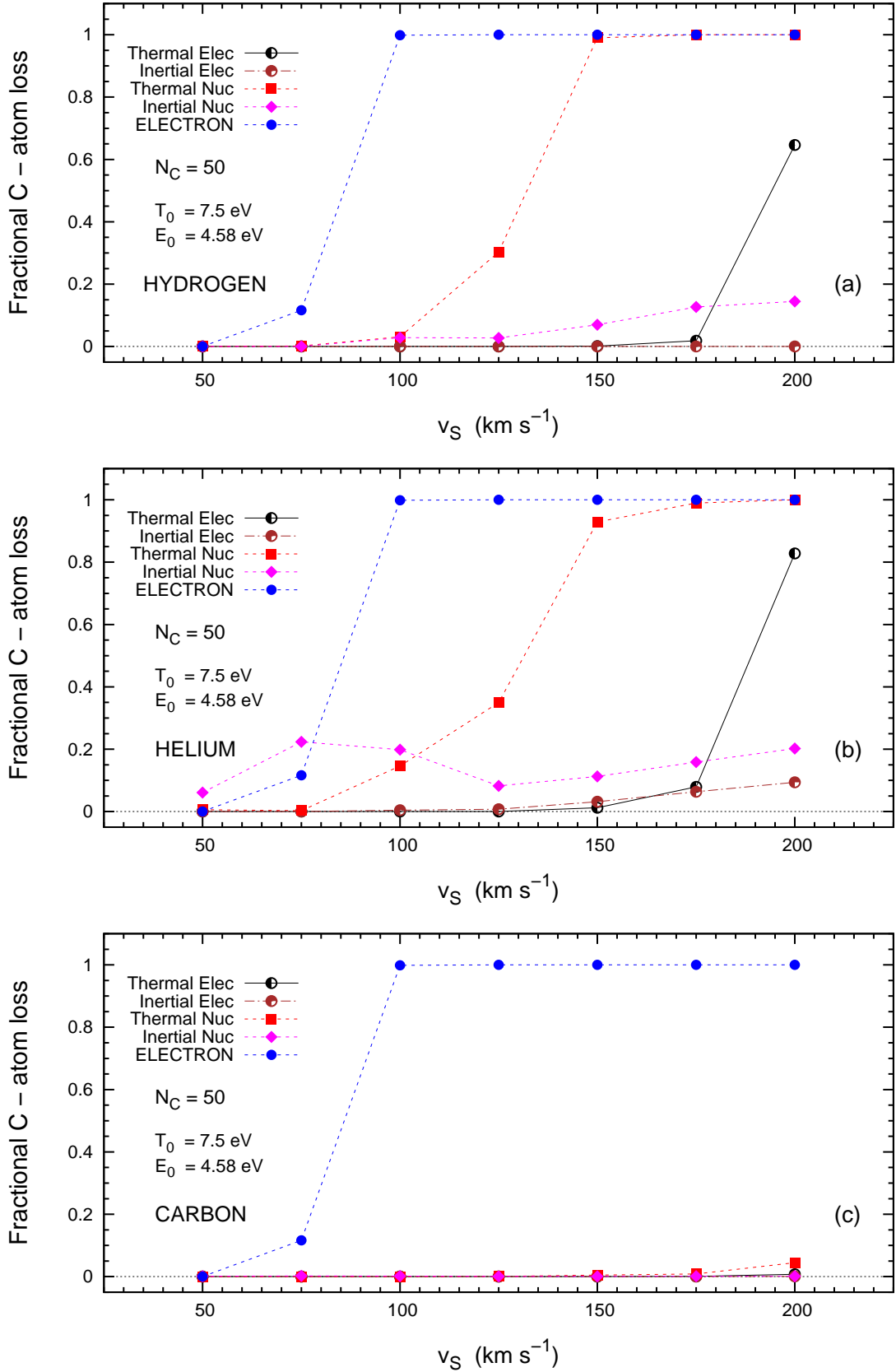}
  \caption{The fractional C-atom loss $F_{\rm L}$, due to collisions
           with H, He and C ions and electrons, as a function of the shock
           velocity. $F_{\rm L}$ is defined as the total number of ejected carbon atoms 
           divided by the initial number $N_{\rm C}$ of C-atoms in the PAH molecule.
           The destructive effect of thermal and inertial `sputtering'
           induced by collisions with the projectiles 
           are shown. In the nuclear case, each lost atom is the result 
           of a single collision with a given projectile, in other words, 
           every collision removes a C-atom from the molecule so, the 
           number of destructive collisions equals the number of ejected
           carbon atoms. For electrons as projectiles and for electronic 
           interaction, each collision leads to the ejection of two 
           carbon atoms. The points are calculated for a 50 C-atoms PAH assuming 
           the nuclear threshold energy $T_{0}$ = 7.5 eV and the electronic
           dissociation energy $E_{0}$ = 4.58 eV. 
           }
  \label{global50_fig}
\end{figure*}

\begin{figure*}
  \centering
  \includegraphics[width=.8\hsize]{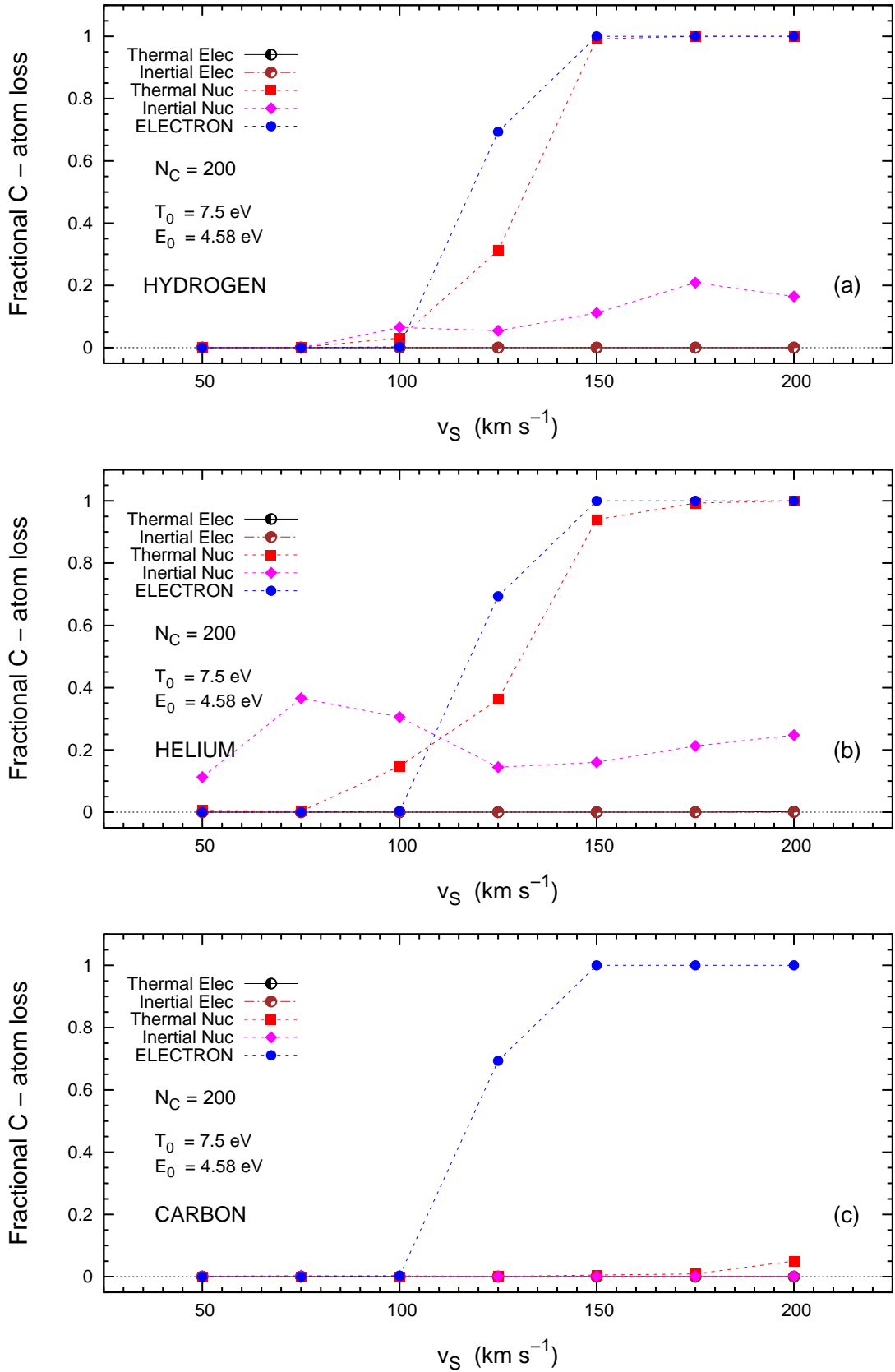}
  \caption{Same as Fig. \ref{global50_fig} calculated for a 200 carbon atoms
           PAH. 
           }
  \label{global200_fig}
\end{figure*}

A summary of our findings from Fig. \ref{global50_fig} and \ref{global200_fig}
is presented in Fig. \ref{fraction_fig}. This shows the fractional atomic loss, 
$F_{\rm L}$, due to electron and ion collisions, calculated for the two PAH sizes 
$N_{\rm C}$ = 50 and 200. To show how the fractional loss changes as a function
of the adopted value for $E_{0}$, we added the results obtained assuming for the 
electronic dissociation energy the values 3.65 and 5.6 eV, lower and higher 
respectively than our standard value 4.58 eV, 

For ionic collisions, $F_{\rm L}$ is determined by nuclear interaction: inertial
for low $v_{\rm S}$ and thermal for high $v_{\rm S}$. For the 50 C-atoms PAH, the
electronic contribution emerges for the lowest value of $E_{0}$, 3.65 eV. As already 
mentioned, the dissociation probability increases for decreasing $E_{0}$, so we are 
not surprised to find $F_{\rm L}$ enhanced by electronic excitation (both inertial and 
thermal). The electronic contribution desappears for higher dissociation energies, as 
demonstrated by the coincidence between the curves for $E_{0}$ = 4.58 and 5.6 eV. 
For $N_{\rm C}$ = 200, all three ionic curves are coincident, indicating that electronic 
excitation by impacting ions does not contribute to PAH destruction below 150 km s$^{-1}$. 
Above this value, the carbon loss due to electronic interaction is covered by the other 
processes. 

The shift between the ionic curves for the two PAH sizes is due to the small 
differences in the velocity profiles -- due to betatron acceleration --, which 
imply in the inertial case a slightly higher damage for the bigger PAH. In the 
thermal case $F_{\rm L}$ is instead independent on the PAH size because the number 
of ejected C-atoms scales linearly with $N_{\rm C}$ (see Sect. 4.1). The resulting 
effect is an almost linear rise for both PAH sizes up to 150 km s$^{-1}$, beyond 
which the destruction is complete.

The `Electron' curves reproduce the behaviour observed in Fig. \ref{global50_fig} 
and \ref{global200_fig}. The comparison with the ionic $F_{\rm L}$ clearly 
indicates that a 50 carbon atoms PAH is already damaged in a non-negligible 
way in low velocity shocks (50 - 75 km s$^{-1}$, ion collisions) and is totally
destroyed above 100 km s$^{-1}$ (electron collisions). When $E_{0}$ = 3.65 eV, the
fraction of ejected C-atom for $v_{\rm S}$ = 75 km s$^{-1}$ increases from 0.1 to 
$\sim$0.35 but the minimum shock velocity for complete destruction is unchanged 
(100 km s$^{-1}$). For $v_{\rm S} \geq$ 75 km s$^{-1}$, the 5.6 eV curve is almost
parallel to the 3.65 eV curve and shifted by 25 km s$^{-1}$ toward higher shock
velocities. For the 200 C-atom PAH, the carbon atom loss is dominated by ionic 
collisions for shocks with velocity below 100 km s$^{-1}$. Above this value, 
the combined effect of ions and electrons leads to a complete destruction. A
lower electronic dissociation energy shifts back by 25 km s$^{-1}$ the minimum
shock velocity required for total carbon ejection. When $E_{0}$ = 5.6 eV, destruction
starts to be important only above 125 km s$^{-1}$, and becomes almost complete at
150 km s$^{-1}$.

We adopt the quantity $F_{\rm L}$ as destruction efficiency to calculate the PAH 
lifetime in shocks.

\subsection{Uncertainties discussion}

The main sources of uncertainties which have to be considered for this study are
related to the adopted shock profiles, to the accuracy of the fitting function for
the ZBL nuclear stopping cross section, and to the choice of an appropriate value
for the nuclear threshold energy $T_{0}$ and for the electronic dissociation energy $E_{0}$. 

The uncertainties related to the adopted shock profiles here are principally due to our
assumption that we can equivalently treat a small, two-dimensional PAH molecule as a small
three-dimensional grain. In the calculation of the PAH velocity profiles through the shocks 
we use the same formalism as for the grains \citep{jones96}, i.e., we assume that the PAH 
behaves as a three-dimensional grain of the same mass. Any uncertainties are then due to 
the inherent differences in the cross section to mass ratios for PAHs and grains.
As mentioned in Sect. 3, once
the PAH cross section is averaged over all possible
orientations, the differences in the PAH and grain cross sections turn out to be only of
the order of 11\%, for a 50 carbon atom PAH, and are therefore rather small compared to 
the other uncertainties that we discuss here.

The accuracy of the ZBL nuclear stopping cross section depends on the accuracy 
of the single analytical function used by \citet{zie85} to calculate the 
interatomic potentials between atoms. This universal function has been compared 
with experimentally determined potentials, with a resulting standard deviation 
between theory and experiment of 5\% \citep{oconn86}. An additional test has
been made comparing the results from the ZBL function with much more complex 
theoretical calculations including more effects. In this case as well the
results agree within few percent \citep[see][ and references therein]{zie85}. 
\clearpage

\clearpage


\begin{figure*}
  \centering
  \includegraphics[width=.8\hsize]{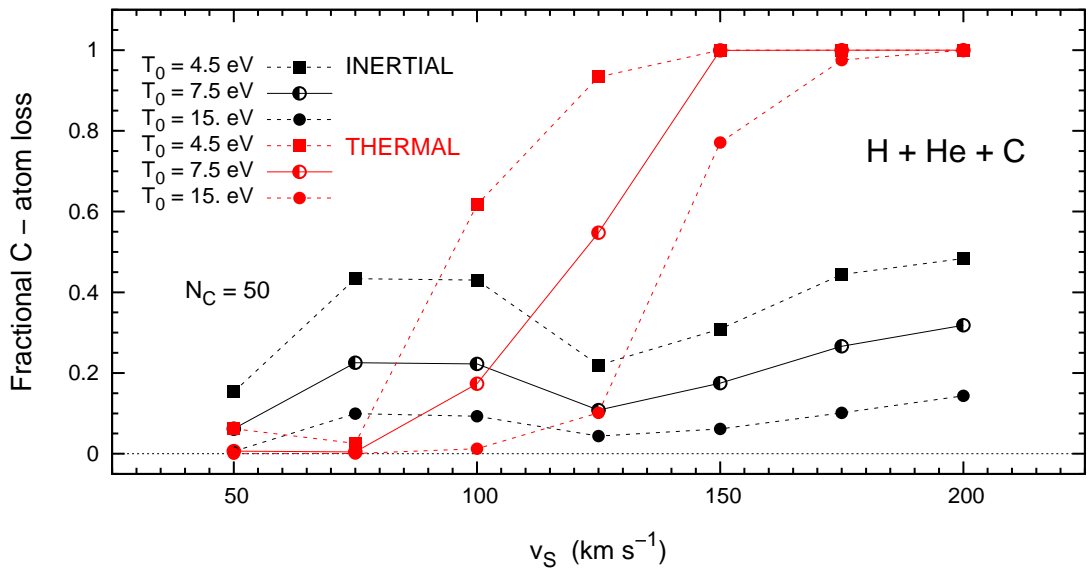}
  \caption{Carbon atom loss due to collisions with thermal and inertial ions
           (H + He + C) via nuclear interaction. The figure shows the comparison 
           betwen the three threshold values $T_{0}$ = 4.5, 7.5 and 15 eV for a 50
           carbon atom PAH. 
           }
  \label{Ncoll3_fig}
\end{figure*}

\clearpage
The main source of uncertainty in the nuclear stopping calculation is the
choice of the threshold energy $T_{0}$. This quantity is not well constrained
(see Sect. 2.2.1) so we explored a set of plausible values. In Fig. \ref{Ncoll3_fig}
we plotted together the values for $F_{\rm L}$ resulting from the total effect of H, 
He and C, calculated in both inertial and thermal case for three different threshold
energies $T_{0}$ = 4.5, 7.5 and 15 eV. The curves corresponding to the highest and 
lowest threshold $T_{0}^{\rm max}$ and $T_{0}^{\rm min}$ identify a region which can 
be interpreted as the variation in the amount of destruction due to the uncertainty 
in the threshold energy. For the inertial case, this uncertainty introduces a variation 
in the destruction efficiency of a factor less than about 2. For the thermal case, the 
uncertainty introduces a shift of the critical shock velocity above which thermal 
destruction is dominant from about 100 to 150 km s$^{-1}$ for $T_{0}$ ranging from 4.5 to 15 eV. 

\begin{figure*}
  \centering
  \includegraphics[width=.8\hsize]{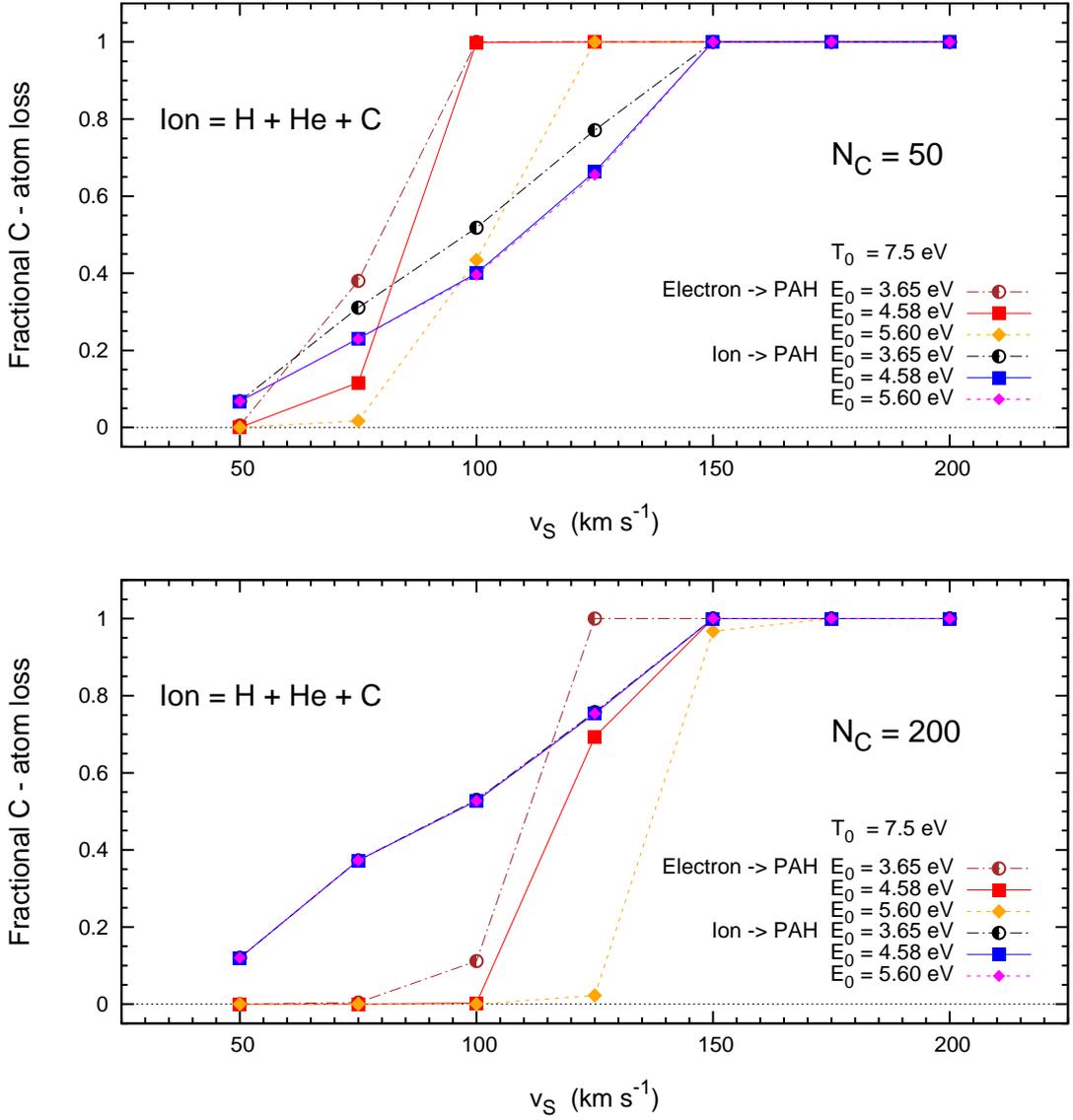}
  \caption{The fractional C-atom loss $F_{\rm L}$ as a function of the shock 
           velocity, calculated for three values of the parameter $E_{0}$: 
           3.65, 4.58 and 5.6 eV. $F_{\rm L}$ is defined as the total number 
           of ejected carbon atoms divided by the initial number $N_{\rm C}$ 
           of C-atoms in the PAH molecule. 
           For the two PAH sizes $N_{\rm C}$ =50 (top panel) and $N_{\rm C}$ = 200
           (bottom panel), the curves labelled `Ion' illustrate the cumulative effect of 
           all destructive processes for the ions considered in this study: 
           inertial and thermal `sputtering' due to 
           nuclear and electronic excitation during PAH collisions with 
           H, He and C in the shocks. The curves labelled `Electron' show
           the destructive effect of thermal electrons. We assume $T_{0}$ = 7.5 eV.
           }
  \label{fraction_fig}
\end{figure*}

\vspace{4pt}
An important issue for the electronic and electron stopping calculation is the 
choice of the value for 
the parameter 
$E_{0}$. We
adopt the value 4.58 eV, which 
has been extrapolated for interstellar conditions from experimental data.
Unfortunately the extrapolation procedure is very model--dependent, so the same
set of experimental data can lead to significantly different values
for the interstellar
$E_{0}$. The problematic fragment binding energy is extensively discussed in MJT, 
the result of a different choice for $E_{0}$ (3.65, 4.58 and 5.6 eV) on PAH 
processing by shocks is shown in Fig. \ref{fraction_fig}.
The differences are quite
significant, indicating the importance of experimental studies on the critical energy 
$E_{0}$ describing the dissociation probability of highly excited PAHs.
To summarize, the errors related to the shock profiles and ZBL fitting function are
quite small, for $T_{0}$ we identified a range of plausible values, but experimental 
determinations would be desirable, while the choice of the parameter $E_{0}$ is very
uncertain and urgently requires a better determination.


\section{Discussion}

\subsection{PAH lifetime in shocks}

To calculate the timescale for supernova shock waves to destroy the interstellar 
PAHs in the Galaxy, $t_{\rm SNR}$, we adopt the same approach used in our
previous works \citep{jones94, jones96}, 
which is based on the method of \citet{mckee89}: 
\begin{equation}\label{timescale1_eq}
  t_{\rm SNR} = \frac{M_{\rm ISM}}{\left(1/\tau^{'}_{\rm SN}\right)\,
           \int \varepsilon(v_{\rm S})\,{\rm d}M_{\rm S}(v_{\rm S})}
\end{equation}
where $M_{\rm ISM}$ = 4.5$\times$10$^{9}\,M_{\sun}$ is the mass of the Galactic 
interstellar medium (gas and dust including PAHs), $\tau^{'}_{\rm SN}$
= 125 yr is the effective interval between supernovae \citep{mckee89}, 
$\varepsilon(v_{\rm S})$ is in this case the efficiency of PAH destruction
by a shock of velocity $v_{\rm S}$ and $M_{\rm S}$ is the mass of gas shocked
to at least $v_{\rm S}$ by a supernova remnant in the Sedov-Taylor stage
(in this stage the energy is conserved, so that $M_{\rm S}\,v_{\rm S}^{2}\propto E$).
In a three-phase model of the interstellar medium \citep{mckee77}, with a ratio
between warm and hot intercloud filling factor of $f_{\rm w}/f_{\rm h}$ = 0.3/0.7 = 0.43,
the mass $M_{\rm S}$ and the timescale $t_{\rm SNR}$ become
\begin{equation}\label{mass_eq}
  M_{\rm S}(v_{\rm S}) = 2914\,M_{\sun}/v_{\rm S7}^{2}
\end{equation}
\begin{equation}\label{timescale2_eq}
  t_{\rm SNR} = \frac{9.7 \times 10^{7}}{\int \varepsilon(v_{\rm S7})
           /v_{\rm S7}^{3}\,{\rm d}v_{\rm S7}}\,yr
\end{equation}
were $v_{\rm S7}$ is shock velocity in units of 100 km s$^{-1}$ and the 
assumed mean supernova energy is 10$^{51}$ erg. 

%
\begin{table*}
\caption{PAH destruction analytical fit parameters and survival timescales
         for electron and ion collisions.
}
\label{fit_tab}
\renewcommand{\footnoterule}{} 
\begin{center}
\begin{tabular}{l c c c c c c}     
\hline\hline
\noalign{\smallskip}
   &  \multicolumn{2}{c}{Electron}  &  \multicolumn{2}{c}{Ion}  &  \multicolumn{2}{c}{Fit}\\
   &   $N_{\rm C}$ = 50   &  $N_{\rm C}$ = 200  &  $N_{\rm C}$ = 50   &  $N_{\rm C}$ = 200   &  \multicolumn{2}{c}{parameters} \\
\noalign{\smallskip}
\hline
\noalign{\smallskip}
$\varepsilon(v_{\rm S7}) = m_{1}x+q_{1}$ & $0.50 \leq v_{\rm S7}\leq 0.75$   &  ---  &  ---  &  ---  &  $m_{1}$ = 0.464  &  $q_{1}$ = - 0.232  \\
$\varepsilon(v_{\rm S7}) = m_{2}x+q_{2}$ & $0.75 \leq v_{\rm S7}\leq 1.00$   &  ---  &  ---  &  ---  &  $m_{2}$ = 3.536  &  $q_{2}$ = - 2.536  \\
$\varepsilon(v_{\rm S7}) = m_{3}x+q_{3}$ &  --- &  --- &  ---  &  $0.50 \leq v_{\rm S7}\leq 1.50$    &  $m_{3}$ = 0.879  &  $q_{3}$ = - 0.319  \\
$\varepsilon(v_{\rm S7}) = a x^{b}$     &   --- &  ---  &  $0.50 \leq v_{\rm S7}\leq 1.50$  &  ---   &  $\;\;a$ = 0.405 &  $b$ = 2.232  \\
$\varepsilon(v_{\rm S7}) = c\,\ln\,(x^{d})\,/x$  &   --- &  $0.50 \leq v_{\rm S7}\leq 1.50$  &   ---  &   ---    & $\;\;c$ = 5.366 &  $d$ = 0.7    \\
$\varepsilon(v_{\rm S7}) =  k$  & $1.00 \leq v_{\rm S7}\leq 2.00$  & $1.50 \leq v_{\rm S7}\leq 2.00$  &  
$1.50 \leq v_{\rm S7}\leq 2.00$  &  $1.50 \leq v_{\rm S7}\leq 2.00$  &  \multicolumn{2}{c}{$k$ = 1}    \\
\noalign{\smallskip}
\hline
\noalign{\smallskip}
\multicolumn{1}{c}{$t_{\rm SNR}$  (yr)}  &  1.6 $\times$10$^{8}$  &  4.0 $\times$10$^{8}$  &  1.8 $\times$10$^{8}$  &  1.4 $\times$10$^{8}$  &    &   \\ 
\noalign{\smallskip}
\hline
\end{tabular}
\end{center}
Note: Fit parameters and survival timescales calculated assuming $T_0$ = 7.5 eV and $E_0$ = 4.6 eV. \\
\end{table*}

Using our calculated fractional destruction data, we derived analytical
expressions for the destruction efficiency $\varepsilon(v_{\rm S7})$ for 
electrons and ions. The ionic term represents the total contribution of  
all considered ions (H, He and C) and processes (nuclear and electronic
stopping, inertial and thermal). Below 150 km s$^{-1}$, we adopt a power 
law fit for $N_{\rm C} = 50$ and a linear fit for $N_{\rm C} = 200$, while 
for the remaining velocities the efficiency is 1. For electrons, 
for $F_{\rm L}<1$, the destruction efficiencies are well fit by two linear 
functions ($N_{\rm C} = 50$) and a logarithmic function ($N_{\rm C} = 200$). 
The analytical fits reproduce the calculated data within few percent. The 
functional form, fitting parameters and corresponding timescales calculated 
from Eq. \ref{timescale2_eq} are reported in Table \ref{fit_tab}.

For electron collisions we find $t_{\rm SNR}$ = 1.6 $\times$10$^{8}$ yr  for 
$N_{\rm C} = 50$ and 4.0 $\times$10$^{8}$ yr for $N_{\rm C} = 200$. In case of 
ion collisions the lifetimes are 1.8 $\times$10$^{8}$ yr and 1.4 $\times$10$^{8}$ yr
for the small and big molecule respectively. The largest uncertainty in these 
lifetimes result from the uncertainty in the values adopted for $T_0$ and $E_0$ 
(cf. Sect. 4.5). If we assume the lowest values considered (4.5 and 3.65 eV), 
these lifetimes decrease to 9.2$\times$10$^{7}$ yr and 8$\times$10$^{7}$ yr
(for $N_{\rm C} = 50$ and 200 respectively), while the maximum values considered for 
these energies result in lifetimes of 2.5$\times$10$^{8}$ yr and 3.3$\times$10$^{8}$ yr. 
for the small and big PAH. Thus, while these parameters are quite uncertain, the 
derived values for the lifetimes are quite robust. Essentially, PAHs are destroyed 
by shocks larger than about 100 km/s and, typically, interstellar gas encounters 
such shocks once every 100 million years \citep{mckee89}. 

From the results obtained assuming our standard values ($T_0$ = 7.5 eV and $E_0$ = 4.6 eV), 
we argue then that small PAHs are preferentially destroyed by electrons whereas big 
PAHs are more affected by ions.  

Our derived values for $t_{\rm SNR}$ for PAHs are significantly shorter than the 6
$\times$10$^{8}$ ~yr calculated by \citet{jones96} for graphite/amorphous carbon 
grains in the warm intercloud medium. Ignoring betatron acceleration, the total 
number of collisions per C-atom required for stopping the inertial motion is 
independent of grain size. The difference in lifetimes reflects then a difference 
in sputtering efficiency. This is not surprising because of the different 
approaches adopted for the ion-particle interactions for PAHs and grains. 
In both cases, and for the shocks that we consider here, it is the sputtering processes 
that completely dominate dust destruction. In the case of grains not every atom that is 
`knocked on' by an incident ion is lost, i.e. sputtered, from the grain. The displaced 
atom is often embedded deeper into the grain and therefore not sputtered from the grain 
even when the displacement energy significantly exceeds the threshold energy for target 
atom displacement. This is reflected in the fact that, for grains, the sputtering yield 
is usually much less that unity. The shorter lifetime for the PAHs is then be ascribed 
to the fact that in any incident ion interaction the target atom is always lost from the 
PAH when the energy to displace it is greater that the required threshold energy. In this 
case the equivalent PAH sputtering yield is then unity.

Our derived PAH lifetime is much closer to the value of $\sim$2$\times$ 10$^{8}$~yr
found for a size distribution of hydrogenated amorphous carbon (a-C:H) grains, typical of
the diffuse interstellar medium, by \citet{serra08}. This is perhaps just coincidental; it 
has its origin in lower average binding energy of C-atoms in amorphous carbon than in 
graphite. a-C:H grains are found, as we find for PAHs, to be more susceptible to 
sputtering erosion than graphite/amorphous carbon grains \citep{jones96}. However 
PAHs are much more susceptible to erosion than a-C:H in fast shocks ($\geq 150$~km~s$^{-1}$) 
and this is simply due to their small sizes. Thermal sputtering in the hot post-shock gas of
fast shocks is proportional to the surface area and, small particles having a larger surface 
area per unit mass than large particles, are more rapidly eroded.

\subsection{Astrophysical implications}

As shown in Sect. 5.1, the PAH lifetime against shock destruction is much shorter
than the stardust injection timescale into the interstellar medium 
$t_{\rm inj}$ = 2.5 $\times$10$^{9}$ yr. Gas shocked to velocities of the order 
of 50 - 150 km s$^{-1}$ is observed in many regions of the interstellar medium: 
e.g. toward the star \object{$\zeta$ Ori} \citep{welty02}, in Herbig-Haro jets in the Orion 
and Vela star forming regions \citep{podio06}, and in the local interstellar cloud 
\citep{slavin08}. Hence, according to our calculations, PAH destruction should be 
widespread in the ISM.

If we assume globally that the same $t_{\rm inj}$ holds for PAHs as for `standard'
interstellar dust, i.e., that dust and PAH formation are coeval in and around 
evolved stars, our calculated survival times for PAHs, i.e., $t_{\rm SNR}$, indicate 
that they need to be re-formed in the ISM even more rapidly than the larger
interstellar grains. If PAHs are formed by the fragmentation of larger carbonaceous 
grains then their `effective' survival time must just be the same as that of the 
larger grains from which they originate. However, this can only be true for low 
velocity shocks, or turbulent regions of the ISM, where grain-grain collisions at 
relatively low velocities (of the order of a few km s$^{-1}$) can form PAHs via 
fragmentation and where there is no associated destructive process in operation.

On a region by region basis our results indicate that PAHs should not exist in 
environments shocked to high velocities ($> 100$ km s$^{-1}$). We conclude that 
PAHs that exist in unshocked regions do not survive the passage of shocks with 
velocities above $100 - 150$ km s$^{-1}$ (depending on their size). They are in 
fact destroyed rather early in in the shock at shocked column densities of the 
order of $10^{16} - 10^{17}$ cm$^{-2}$. Any `daughter' PAHs produced in the 
post-shock region, by grain fragmentation in grain-grain collisions at shocked 
column densities of the order of $10^{17} - 10^{18}$ cm$^{-2}$, will be destroyed 
by erosion due to their high injection velocities into the gas following the 
fragmentation of their larger, 'parent' grains that undergo betatron-acceleration
\citep{jones96}. Thus, high velocity shocks destroy all the PAHs that they both 
interact with and produce by fragmentation in high velocity grain-grain collisions. 

In contrast observations show that PAHs lock up about 3\%\ of the elemental carbon in 
the ISM \citep[cf.][]{tielens08}.
Disentangling these two scenarios is not easy observationally. 
Here we now consider the case where the PAH emission is assumed to come from within the 
shocked region. This scenario requires an efficient (re)formation route for PAHs in the 
diffuse ISM (see above). However, this is difficult to understand since PAHs are a product of high 
temperature chemistry involving abundant carbon bearing precursors such as CH$_{4}$ and 
C$_{2}$H$_{2}$. In the low temperature diffuse ISM, an O-rich environment, these precursor 
species are never really very abundant. This conundrum is very reminiscent of the general 
problem of rapid dust destruction in the ISM and the long injection time scale for freshly 
formed dust \citep{dwek80, draine79, jones94, jones96}.
We note that, while dust grains may be rapidly covered by (thin)protective coatings between 
successive shock passages \citep[cf.][]{tielens98}, this is not a way out of this conundrum 
for PAHs. Perhaps, PAHs can be formed through prolonged photolysis of ice mantles accreted 
inside dense molecular clouds, although, presently, there is no experimental support for this 
suggestion. Lastly, observations have revealed the presence of PAHs associated with hot shocked 
gas in stellar (eg., \object{M17}) and galactic (e.g., M82) wind regions (cf. Sect. 1). As will be discussed 
more extensively in MJT, these PAHs likely trace entrained cold gas which has not been fully 
exposed to the destructive effects of high velocity shocks.


\begin{figure*}
  \centering
  \includegraphics[width=.8\hsize]{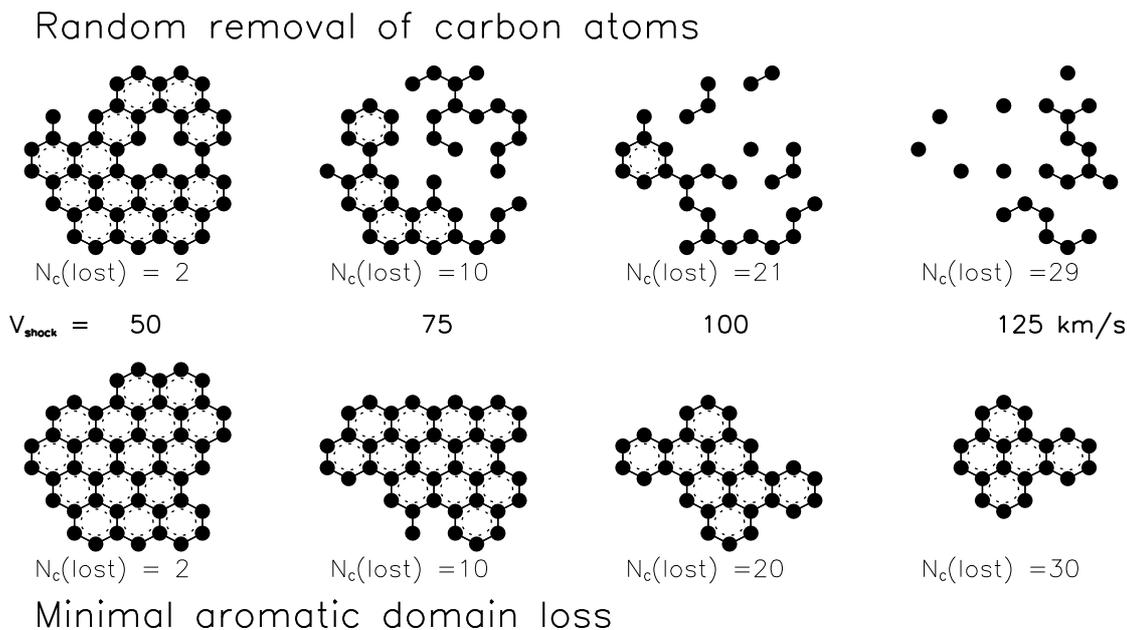}
  \caption{The evolution of a 50 carbon atom PAH following the loss of $N_{\rm C}$(lost) 
           carbon atoms, as a function of the shock velocity,
           for the two limiting cases: instantaneous and random removal of 
           the lost carbon atoms (top row), and carbon atom removal only from 
           the periphery of the molecule (bottom row).
           }
  \label{PAHevol_fig}
\end{figure*}


For $v_{\rm S} < $ 75 km s$^{-1}$, PAH are disrupted only by inertial `sputtering' due
to nuclear interactions. \citet{serra08} have studied the erosion of hydrogenated
amorphous carbon (a-C:H) arising from ion irradiation in shocks in the warm intercloud
medium, using exactly the same shock profiles that we use for our PAH study, and it is
interesting to note that, for a 50 carbon atom PAH molecule, the percentage of destruction
is the same as for a-C:H, indicating that the lower density, more easily sputtered a-C:H
and PAHs exhibit similar erosion characteristics in lower velocity shocks
($\simeq 125$ km~s$^{-1}$), as discussed in the previous section. 

We now consider what happens to a PAH as a result of the ejection of aromatic
carbon atoms, by the impacting ions, as a function of the fraction of carbon
atoms removed from the structure. Our results show that ionic collisions severely 
modify PAH in shocks with velocities between 75 - 150 km s$^{-1}$. Nuclear and 
electronic interaction lead to distinctly different molecular destruction routes. 
Specifically, electronic excitation (either by impacting ions or electrons) lead 
to a high vibrational excitation of the PAH and this PAH will relax by `losing' 
the weakest link in its skeleton. Initially, this will be the peripheral H-atoms 
or other functional groups. For large PAHs (50-100 C-atoms), the resulting `pure-C' 
may quickly isomerize to very stable carbon cluster such as fullerenes. Fullerenes 
are very stable against C$_2$ loss. Measured dissociation energies of fullerenes are 
in the range of 7 to 9.5 eV, with C$_{60}$ itself at $E_{\rm 0}=9.8 \, \pm \, 0.1$ eV 
\citep{tomita01}.

In contrast, nuclear interaction will act on the C-atom hit by the projectile ion. 
This will randomly remove C-atoms from the C-skeleton of the PAH. Some isomerization 
may occur if internal excitation energy is left behind. However, unless the C-atom 
loss in the shock is very large, likely this is insufficient to affect the overall 
PAH structure. In a simple `geometrical' analysis, if one randomly removes C atoms 
from a PAH with $N_{\rm C} = 50$, in the absence of any annealing of the PAH chemical 
structure, we find that the loss of 10\% of the C atoms leads to a loss of of the 
order of $\approx 50$\,\%  of the aromatic character. However, the loss of $\approx 20\,\%$ 
of the C atoms leads to the almost complete loss of aromatic character and to the
onset of the fragmentation of the molecule. 
Fig. \ref{PAHevol_fig} illustrates these effects and shows the PAH evolution following 
the loss of carbon atoms, $N_{\rm C}$(lost), for the two limiting cases: 1) where there is 
an instantaneous and random removal of the lost carbon atoms (appropriate for nuclear 
interaction) and 2) where the carbon atoms are removed only from the periphery in order 
to preserve aromatic domain as much as possible (likely appropriate for electronic 
excitation through either impact ions or electrons). The reality of PAH erosion in shocks 
may well lie somewhere between these two extremes and may also involve isomerisation 
and the formation of five-fold carbon rings that distort the structure from a perfectly 
two-dimensional form. This then begs the question as to the exact form and structure 
of small carbon species once growth resumes by atom insertion and addition.

After the shock wave has passed, the resulting PAH can react chemically with impacting 
H, C, O, and N atoms. A distinction can be made between H and O in the one hand and C 
and N on the other hand, The latter can restore the stable aromatic character of the 
PAH skeleton while the former lead only to very unstable structures. Likely, then, H 
and O atom addition can be reversed by UV photon absorption without loss of C 
\citep{allamandola89}. Hence, the PAH carbon skeleton may be able to `repair' itself 
to an aromatic structure, possibly incorporating N-atoms. The possible presence of 
N-atoms deeply in the C-skeleton of PAHs in the diffuse medium (but not in C-dust 
birth sites such as C-rich planetary nebulae) has been inferred from the peak 
position of the 6.2 $\mu$m band \citep{peeters02}. It has also been suggested that 
the $6.2\,\mu$m band position variations can be explained by a varying aliphatic to 
aromatic carbon content in carbonaceous particles \citep{pino08}. The difference in 
molecular structure between circumstellar PAHs and the general ISM implies an 
active chemistry in the ISM that is able to insert N atoms deeply in the carbon 
skeleton. This is very puzzling given the very stable character of the aromatic 
backbone and the low temperature of the ISM. Here, we surmise that, unlike UV 
photolysis or chemical attack, nuclear interactions in interstellar shocks 
may be a viable pathway to promote N-incorporations deep inside PAHs. However, 
given that the nitrogen abundance is a factor of a few lower than that for carbon -- 
depending on the fraction of carbon remaining in dust and PAHs -- carbon atom 
insertion ought to be favoured over that for nitrogen, in the absence of any 
chemically-selective route for nitrogen insertion.

\section{Conclusions}
 
We have extensively studied the effects of PAH processing by shocks with velocities 
between 50 and 200 km s$^{-1}$, in terms of collisions with ions and electrons which
can lead to carbon atom loss, with a consequent disruption and destruction of the
molecule.
 
An ionic collision consists of two simultaneous processes which can be treated
separately: a binary collisions between the projectile and one single atom in the 
target (nuclear interaction) and the energy loss to the atomic electrons (electronic 
interaction). For the nuclear interaction, we modified the existing theory in order 
to treat collisions able to transfer energy \textit{above} a specific threshold. 
This is the case we are interested in, which has not been treated in previous studies. 
For electronic interaction and collisions with electrons we developed specific models 
for PAH, described in MJT.
 
The PAH dynamics in the shocks is evaluated using the same approach as
in our previous work \citep{jones94, jones96}. For nuclear
interaction, the level of carbon atom loss increases for decreasing
values of the threshold energy $T_{0}$. We adopt $T_{0}$ = 7.5 eV as a
reasonable value, but experimental determinations are necessary.
In ionic collisions, the carbon contribution to PAH destruction is totally 
negligible because of its very low abundance with respect to H and He.
The fractional destruction induced by nuclear excitation increases with the
PAH size, while in case of electronic excitation and electron collisions is lower 
for higher $N_{\rm C}$ values, i.e. bigger PAHs are more resistant than the smaller 
ones against electron and electronic processing.
 
The parameter $E_{0}$ required for the evaluation of PAH destruction due to 
electron and electronic interaction is unfortunately not well constrained. We adopt
a value of 4.58 eV consistent with extrapolations to interstellar conditions, but
better determinations would be desirable.
    
Electronic interaction, both inertial and thermal, plays a marginal
role in PAH processing by shocks. We find that 50 carbon atoms PAHs
are significantly disrupted in ionic collisions for shock velocities
below 75 km s$^{-1}$, mainly by inertial `sputtering' by helium during
nuclear interaction. Our results indicate $5-15\,\% $ C atom loss,
sufficient to cause a severe de-naturation of the PAH aromatic
structure. Above 100~km~s$^{-1}$ such PAHs are instead totally
destroyed by collisions with thermal electrons. For $N_{\rm C} = 200$,
PAHs experience increasing damaging caused by nuclear ionic collisions
up to 100~km~s$^{-1}$, which turns into complete atomic loss for
higher velocities.  In this case the destruction is due to the
combined effect of electrons and nuclear interaction with thermal
ions.

The calculated PAH lifetime against destruction, $t_{\rm SNR}$, is 
1.6 $\times$10$^{8}$ yr and 1.4 $\times$10$^{8}$ yr for $N_{\rm C}$ = 50 and 
200 respectively. Small PAHs are preferentially destroyed by electrons, big PAHs by 
ions. The calculated lifetimes are smaller than the values found for carbonaceous 
grains (6 $\times$10$^{8}$ yr) but close to that for hydrogenated amorphous 
carbon ($~$2 $\times$10$^{8}$ yr), and far from the stardust injection 
timescale of 2.5 $\times$10$^{9}$ yr. The presence of PAHs in shocked regions 
therefore requires an efficient reformation mechanism and/or a protective environment.

We surmise that the molecular structure of PAHs is strongly affected by shock 
processing in the interstellar medium. Electronic excitation by impacting ions or 
electrons may lead to isomerization into stable pure-C species such as fullerenes. 
In contrast, nuclear interaction may lead to the formation of N-containing PAHs. 
Further laboratory studies are required to demonstrate the viability of these 
chemical routes.

\appendix

\section{$S_{\rm n}$, $\sigma$ and $\langle T \rangle$}

A fundamental quantity to describe the nuclear scattering is the \textit{energy 
transfer cross section} $\sigma(E,T)$, which is function of the kinetic energy 
$E$ of the projectile and of the energy $T$ transferred to the target by the 
projectile in a single collision. At low energies ($\varepsilon \lesssim$ 1), 
an approximated expression for the cross section can be calculated using the 
power approximation to the Thomas-Fermi model of interatomic interaction, i.e. 
with a potential of the form $V(r) \propto r^{-1/m}$, where $r$ is the distance 
between colliding nuclei and $m$ is a parameter related to the steepness of the 
interatomic potential. The quantity $m$ can also be interpreted as an indicator 
of the energy of the projectile, varying slowly from $m$ = 1 at high energies to 
$m \approx$ 0 at low energies \citep{lind68, winter70}. We have that

\begin{equation}\label{dsigma_eq}
 {\rm d} \sigma(E,T) \cong C_{m}\,E^{-m}\,T^{-1-m}\,{\rm d} T\;\;\;\;\;
              0\lesssim T \lesssim T_{\rm m}
\end{equation}
\smallskip
\noindent
with

\begin{equation}\label{cm_eq}
 C_{m} = \frac{\pi}{2}\:\lambda_{m}\:a^{2}\,\left(\frac{M_{1}}{M_{2}}\right)^{m}\,
\left(\frac{2\,Z_{1}\,Z_{2}\,e^{2}}{a}\right)^{2m}
\end{equation}
\smallskip
\noindent
where $T_{\rm m}$ is the maximum transferable energy, corresponding to a
head-on collision (impact parameter $p$ = 0). The dimensionless
quantity $\lambda_{m} = \lambda_{m}(m)$ varies slowly from 0.5 for
$m$~=~1 \citep[high energy, i.e. pure Rutherford scattering,]
[]{simm65}, to 24 for $m$ = 0 (very low energy).

The power approximation of the Thomas-Fermi cross section 
(Eq. \ref{dsigma_eq}) gives the following expression for the 
nuclear stopping cross section, obtained evaluating the integral 
in Eq.~\ref{dEdR_eq} between $0$ and $T_{\rm m}$:
\begin{equation}\label{Sn_eq}
  S_{\rm n}(E) = \frac{1}{1-m}\:C_{m}\:\gamma^{1-m}\:E^{1-2m}
\end{equation}
\smallskip
\noindent
For Thomas-Fermi interaction, using Eq. \ref{eps_eq}, \ref{cm_eq},
\ref{tm_eq} and \ref{Sn_eq}, this leads to
\begin{equation}\label{Sn_gen_eq}
 S_{\rm n}(E) = 4\:\pi\:a\:Z_{1}\:Z_{2}\:e^{2}\:
               \frac{M_{1}}{M_{1}+M_{2}}\;s_{\rm n}(\varepsilon)
\end{equation}
\smallskip
\noindent
with
\begin{equation}\label{sn_red_eq}
 s_{\rm n}(\varepsilon) = \frac{\lambda_{m}}{2\,(1-m)}\;\varepsilon^{1-2m}
\end{equation}
\smallskip
\noindent
For heavy screening ($\varepsilon \ll$ 1, $m \cong$ 0) the
accuracy of the Thomas-Fermi reduced stopping cross section 
$s_{\rm n}(\varepsilon)$ to reproduce the experimental data is at
best a factor of two. For this reason we adopt instead the 
Universal reduced stopping cross section $s_{\rm n}^{\rm U}$ 
(Eq. \ref{sn_ZBL_eq}) from \citet{zie85} with the appropriate
screening length $a_{\rm U}$ (Eq. \ref{aU_eq}), which provides 
a good fit to the experimental data at low energies as well.

The nuclear stopping cross section above threshold is given by
Eq. \ref{SnTH_eq}. Remembering that $T_{\rm m} = \gamma E$ and 
$T_{0} = \gamma E_{\rm 0n}$, this can be rewritten as follows

\begin{eqnarray*}
  S_{\rm n}(E) & = & \frac{C_{m}\,E^{-m}}{1-m} \, [T_{\rm m}^{1-m} - T_{0}^{1-m}] \\[.3cm]
              & = & \frac{C_{m}\,E^{-m}}{1-m} \,T_{\rm m}^{1-m}\,-\, 
                    \frac{C_{m}\,E^{-m}}{1-m} \, T_{0}^{1-m} \\[.3cm]
              & = & \frac{C_{m}\,\gamma^{1-m}}{1-m} \,E^{1-2m}\,-\,
                    \frac{C_{m}\,\gamma^{1-m}}{1-m} \, E^{-m} \, E_{\rm 0n}^{1-m}\\
\end{eqnarray*}
\noindent
The first term in the right side of the equation is equal to the 
nuclear stopping cross section in the no-threshold case ($T_{0}$~=~0)
\begin{eqnarray*}
  \frac{C_{m}\,\gamma^{1-m}}{1-m} \,E^{1-2m} & = & 
  4\:\pi\:a\:Z_{1}\:Z_{2}\:e^{2}\:\frac{M_{1}}{M_{1}+M_{2}}\;s_{\rm n}(\varepsilon)
   \; \equiv \;  S^{0}_{\rm n}(E).  \\
\end{eqnarray*}
\noindent
In the second term we use the equality $E^{-m} = (E^{1-2m}/E^{1-m})$
to obtain
\begin{eqnarray*}
  \frac{C_{m}\,\gamma^{1-m}}{1-m} \, E^{-m} \, E_{\rm 0n}^{1-m} & = & 
  \frac{C_{m}\,\gamma^{1-m}}{1-m} \, E^{1-2m} \, \left(\frac{E_{\rm 0n}}{E}\right)^{1-m} \\[.3cm]
        & = & 4\:\pi\:a\:Z_{1}\:Z_{2}\:e^{2}\:\frac{M_{1}}{M_{1}+M_{2}}\;s_{\rm n}(\varepsilon) \times 
              \left(\frac{E_{\rm 0n}}{E}\right)^{1-m}\\    
\end{eqnarray*}
\noindent
Combining the two we obtain the following expression for $S_{\rm n}$
\begin{eqnarray*}
  S_{\rm n}(E) & = & 4 \pi a Z_{1} Z_{2}\, e^{2} \frac{M_{1}}{M_{1}+M_{2}} s_{\rm n}(\varepsilon)
  \left[1 - \left(\frac{E_{\rm 0n}}{E}\right)^{1-m}\right] \\
\end{eqnarray*}
\noindent
The total cross section is given by Eq. \ref{sigmaTH_eq}
\begin{eqnarray*}
  \sigma(E) & = & \frac{C_{m}\,E^{-m}}{m} \, [T_{0}^{-m} - T_{\rm m}^{-m}] \\[.3cm]
            & = & C_{m}\,E^{-m}\,\frac{T_{0}^{-m}}{m}\,-\,
                  C_{m}\,E^{-m}\,\frac{T_{\rm m}^{-m}}{m}. \\
\end{eqnarray*}
The term $(T_{0}^{-m}/m)$ can be rewritten as follows
\begin{eqnarray*}
  \frac{T_{0}^{-m}}{m} & = & \frac{T_{\rm m}^{1-m}}{1-m} \times \frac{1-m}{m} \times 
                            \frac{T_{0}^{-m}}{T_{\rm m}^{1-m}} \\[.3cm]
       & = &  \frac{\gamma^{1-m}\,E^{1-m}}{1-m} \times \frac{1-m}{m} \times
              \frac{\gamma^{-m}\,E_{\rm 0n}^{-m}}{\gamma^{1-m}\,E^{1-m}} \\[.3cm]
       & = &  \frac{\gamma^{1-m}\,E^{1-m}}{1-m} \times \frac{1-m}{m} \times
              \frac{1}{\gamma\,E}\:\left(\frac{E_{\rm 0n}}{E}\right)^{-m}. \\
\end{eqnarray*}
Then we have
\begin{eqnarray*}
C_{m}\,E^{-m}\,\frac{T_{0}^{-m}}{m} & = & \frac{C_{m}\,\gamma^{1-m}}{1-m} \,E^{1-2m}
                    \times \frac{1-m}{m} \times
                    \frac{1}{\gamma\,E}\:\left(\frac{E_{\rm 0n}}{E}\right)^{-m} \\[.3cm]
              & = & S^{0}_{\rm n}(E) \times \frac{1-m}{m} \times
                    \frac{1}{\gamma\,E}\:\left(\frac{E_{\rm 0n}}{E}\right)^{-m}. \\
\end{eqnarray*}
Using a similar approach we can write the term $(T_{\rm m}^{-m}/m)$ as
\begin{eqnarray*}
  \frac{T_{\rm m}^{-m}}{m} & = & \frac{T_{\rm m}^{-m}}{m} \times \frac{1-m}{1-m} \times
                            \frac{T_{\rm m}}{T_{\rm m}} \\[.3cm]
                      & = & \frac{T_{\rm m}^{1-m}}{1-m} \times \frac{1-m}{m} \times
                            \frac{1}{T_{\rm m}} \\[.3cm]
                      & = & \frac{\gamma^{1-m}\,E^{1-m}}{1-m} \times \frac{1-m}{m} \times
                            \frac{1}{\gamma\,E}. \\
\end{eqnarray*}
Then second term in the right side of the equation then becomes
\begin{eqnarray*}
  C_{m}\,E^{-m}\,\frac{T_{\rm m}^{-m}}{m} & = & \frac{C_{m}\,\gamma^{1-m}}{1-m} \,E^{1-2m}
                     \times \frac{1-m}{m} \times \frac{1}{\gamma\,E} \\[.3cm]
               & = &  S^{0}_{\rm n}(E) \times \frac{1-m}{m} \times \frac{1}{\gamma\,E}. \\
\end{eqnarray*}
Combining the two terms we obtain for $\sigma(E)$ the final expression
\begin{eqnarray*}
  \sigma(E) & = &  4 \pi a Z_{1} Z_{2} e^{2} \frac{M_{1}}{M_{1}+M_{2}} s_{\rm n}(\varepsilon)\,
  \frac{1-m}{m}\frac{1}{\gamma\,E}\,
  \left[\left(\frac{E_{\rm 0n}}{E}\right)^{-m} - 1\right]
\end{eqnarray*}

The expression for the average transferred energy $\langle T(E) \rangle$ derives
directly from Eq. \ref{EtransfTH_eq} when applying the relations  
$T_{\rm m}~=~\gamma E$ and $T_{0} = \gamma E_{\rm 0n}$
\begin{eqnarray*}
  \langle T(E) \rangle & = & \frac{m}{1-m}\, \frac{T_{\rm m}^{1-m} - T_{0}^{1-m}}{T_{0}^{-m} - T_{\rm m}^{-m}} \\[.3cm]
                   & = & \frac{m}{1-m}\, \frac{\gamma^{1-m}\,\left(E^{1-m} - E_{\rm 0n}^{1-m}\right)}
                                        {\gamma^{-m}\,\left(E_{\rm 0n}^{-m} - E^{-m}\right)} \\[.3cm]
             & = & \frac{m}{1-m}\,\gamma \,\frac{E^{1-m} - E_{\rm 0n}^{1-m}}{E_{\rm 0n}^{-m} - E^{-m}}. \\
\end{eqnarray*}

\section{Low and high energy regime above threshold}

In the low energy regime $m$ can be taken equal to 0, so the 
differential cross section d$\sigma$ becomes
\begin{eqnarray}
  {\rm d} \sigma(E,T) & = & \cong \frac{\pi}{2}\,\lambda_{0}\,a^{2}\,T^{-1}\,{\rm d} T.
\end{eqnarray}
The corresponding expressions for $S_{\rm n}(E)$, $\sigma(E)$ and $\langle T(E) \rangle$
above threshold are calculated from Eq. \ref{SnTH_def_eq}, Eq. \ref{sigmaTH_def_eq}
and Eq. \ref{EtransfTH_def_eq} respectively
\begin{eqnarray}
  S_{\rm n}(E) & = & \frac{\pi}{2}\,\lambda_{0}\,a^{2} \,\gamma\, (E - E_{\rm 0n}) \\[.3cm]
   \sigma~(E) & = & \frac{\pi}{2}\,\lambda_{0}\,a^{2}  \,\ln\,\frac{E}{E_{\rm 0n}} \\[.3cm]
       \langle T(E) \rangle & = & \gamma\,(E - E_{\rm 0n})\,\left(\ln\,\frac{E}{E_{\rm 0n}}\right)^{-1}
\end{eqnarray}
\noindent
In the high enery regime $m$ = 1, then we have
\begin{eqnarray}
 {\rm d} \sigma(E,T) \cong \frac{M_{1}}{M_{2}}\,Z_{1}^{2}\,Z_{2}^{2}\,e^{4}\,\pi\,\frac{1}{E}\,T^{-2}\,{\rm d} T
\end{eqnarray}
and consequently
\begin{eqnarray}
   S_{\rm n}(E) & = & \frac{M_{1}}{M_{2}}\,Z_{1}^{2}\,Z_{2}^{2}\,e^{4}\,
   \pi\,\frac{1}{E}\,\ln\,\frac{E}{E_{\rm 0n}} \\[.3cm]
 \sigma~(E) & = & \frac{M_{1}}{M_{2}}\,Z_{1}^{2}\,Z_{2}^{2}\,e^{4}\,
 \pi\,\frac{1}{E}\,\frac{1}{\gamma E_{\rm 0n}} \\[.3cm]
 \langle T(E) \rangle & = & \gamma\,E_{\rm 0n}\,\ln\,\frac{E}{E_{\rm 0n}} \\
\end{eqnarray}

\section{Orientation correction}

To calculate the orientation correction factor $\langle S \rangle$, let us consider 
the two versors $\vec{u}$ and $\vec{n}$, perpendicular to the PAH surface
and forming between each other the angle $\vartheta$. This configuration 
defines two possible orientations for the molecule. The average orientation
is then given by the following integral:
\begin{equation}\label{avOrientation1_eq}
  \langle S \rangle = \frac{1}{2\pi}\,\int \vec{u} \cdot \vec{n}\;{\rm d}\,\Omega
\end{equation}
where ${\rm d}\Omega = \sin\vartheta\,{\rm d}\vartheta\,{\rm d}\varphi$ is the solid angle
element and $\vec{u}~\cdot~\vec{n}$~=~$\cos\vartheta$ is the scalar product
between the two versors. With these substitutions we obtain
\begin{equation}\label{avOrientation2_eq}
  \langle S \rangle = \frac{1}{2\pi}\,\int_{\varphi = 0}^{2\pi}{\rm d}\varphi
        \int_{\vartheta = 0}^{\pi/2}\cos\vartheta\,\sin\vartheta\,{\rm d} \vartheta\;
        =\;\frac{1}{2}
\end{equation}
\smallskip
\smallskip        

\begin{acknowledgements}
We are grateful to L. Allamandola and L. Verstraete for useful 
discussions, and we acknowledge our referee Tom Hartquist for careful reading
and helpful comments. E.R.M. thanks G. Lavaux for support and technical 
assistance and acknowledges financial support by the EARA 
Training Network (EU grant MEST-CT-2004-504604).
\end{acknowledgements}


\end{document}